\documentclass[aps,preprint]{revtex4-1}
\usepackage[utf8]{inputenc}
\usepackage[T1]{fontenc}
\usepackage{amsmath}
\usepackage{amsfonts}
\usepackage{amssymb}
\usepackage{graphicx}
\usepackage{lineno}
\usepackage{tabularx,ragged2e,booktabs}
\usepackage{array}
\usepackage{makecell}

\newcolumntype{C}{>{\Centering\arraybackslash\hspace{0pt}}X}
\newcolumntype{s}{>{\hsize=.5\hsize}\Centering\arraybackslash\hspace{0pt}X}

\begin{document}

\title{Extinction time distributions of populations and genotypes}

\author{David Kessler and Nadav M. Shnerb  }

\affiliation{Department of Physics, Bar-Ilan University,
Ramat-Gan IL52900, Israel.}
\begin{abstract}
\noindent In the long run, the eventual extinction of any biological population is an inevitable outcome. While extensive research has focused on the average time it takes for a population to go extinct under various circumstances, there has been limited exploration of the distributions of extinction times and the likelihood of significant fluctuations. Recently, Hathcock and Strogatz~\cite{hathcock2022asymptotic} identified Gumbel statistics as a universal asymptotic distribution for extinction-prone dynamics in a stable environment. In this study, we aim to provide a comprehensive survey of this problem by examining a range of plausible scenarios, including extinction-prone, marginal (neutral), and stable dynamics. We consider the influence of demographic stochasticity, which arises from the inherent randomness of the birth-death process, as well as cases where stochasticity originates from the more pronounced effect of random environmental variations. Our work proposes several generic criteria that can be used for the classification of experimental and empirical systems, thereby enhancing our ability to discern the mechanisms governing extinction dynamics. By employing these criteria, we can improve our understanding of the underlying mechanisms driving extinction processes.
\end{abstract}
\maketitle

\section{Introduction}

Biological populations are inevitably destined for extinction. Over $99 \%$ of all known species that have ever existed on Earth have already become extinct, and the others are awaiting their inevitable turn. The concern over the anthropogenic acceleration of extinction rates has sparked heated debates in the past decade regarding whether such acceleration is indeed observed in local populations~\cite{dornelas2014assemblage,gonzalez2016estimating} and, if so, what are the global implications of this change. Understanding the likelihood of extinction under specific conditions and the distribution of extinction times is crucial for predicting future extinction events and assessing the threat to biodiversity. The same questions also arise when the objective is to eliminate a particular biological entity, such as in the case of pest control, pathogen eradication, or combating genetic diseases.

The dynamics of biological populations is influenced by deterministic and stochastic factors. At the deterministic level, the dynamics can be classified into two main types: those attracted to a manifold (such as a fixed point) with finite population, and those attracted to an extinction point. In the latter case the population decays over time towards zero. Persistent populations of the first type would be expected to survive indefinitely, while populations of the second type disappear. In the common case of exponential decline, the extinction time is logarithmic in the size of the original population.

Stochasticity makes this picture much more subtle. Since the state of zero population is an absorbing state, the ultimate fate of \emph{any} stochastic dynamics is extinction. The sharp distinction between extinction-prone and stable populations thus blurs, and the focus must switch to the characteristics of the extinction process, and in particular to the statistical properties of extinction times.

Typically, stochasticity in biological systems is quite strong, even under extremely stable experimental conditions~\cite{hekstra2012contingency}. Stochastic fluctuations are usually classified into two categories,  \emph{demographic stochasticity} (or genetic drift or internal noise) and \emph{temporal environmental stochasticity} (extrinsic noise)~\cite{lande2003stochastic}.  Demographic noise reflects the inherent randomness of the birth-death process caused by small-scale random events that affect the reproductive success of individuals in an uncorrelated manner. Temporal environmental stochasticity (TES) is associated with large-scale events that affect entire populations. Mathematically speaking, this implies that the  parameters of a given model (usually, the growth rates) vary in time, where the amplitude and correlation times of the fluctuations characterize the environment. Abundance variations induced by TES are usually proportional to the population size, whereas those induced by demographic stochasticity scale with the square root of population size.  Therefore, demographic stochasticity is typically negligible when population size is large~\cite{kalyuzhny2014niche,kalyuzhny2014temporal,
chisholm2014temporal,bergland2014genomic,kalyuzhny2015neutral,grilli2020macroecological}, but it becomes important at the brink of extinction, or during invasion~\cite{pechenik1999interfacial,dornic2005integration,assaf2017wkb,pande2022quantifying}.

This brings us to a third type of systems: those in which deterministic dynamics is weak or negligible, and stochasticity is the main, or only, driver of fluctuations. In this case, we are talking about neutral dynamics, a topic of great importance in population genetics and community ecology~\cite{kimura1985neutral,Hubbell2001unifiedNeutral,maritan1,kalyuzhny2015neutral}. In sum, our classification contains six types of systems: persistent, extinction-prone, and neutral, each of which can be analyzed under pure demographic noise or under a combination of demographic and environmental stochasticity.

In recent works, Strogatz and Hathcock~\cite{hathcock2019fitness,hathcock2022asymptotic}  analyzed the distribution of extinction times for an extinction prone (negative growth rate, exponentially decaying) population with pure demographic stochasticity. These authors found a universal asymptotic behavior, i.e., that the fluctuations around the expected extinction time obey a Gumbel distribution. Furthermore, the width of this distribution is extremely narrow: while the deterministic mean time to extinction scales with the logarithm of the initial population size $N_0$, the width is $N_0$-independent. Therefore, relative fluctuations around the mean vanish as $N \to \infty$.

Here we would like to extend the work of~\citet{hathcock2022asymptotic} and to consider statistics of extinction times in all the six typical scenarios mentioned above.  Some of these cases have already been discussed in the literature (see details below), but we believe that there would be great benefit in presenting them side by side so that a researcher interested in this topic can see the different alternatives. Furthermore, to the best of our knowledge, the results regarding neutral systems are original. In what follows we devote a single section to each of the six scenarios. In the last section we will discuss the results and provide a general outlook.

\section{Extinction-prone dynamics in a fixed environment} \label{sec2}

In this section we first revisit the class of systems considered by~\citet{hathcock2022asymptotic}, for which the Gumbel distribution is a universal limit. In the next subsection we provide an example of a ``non-Gumbel" scenario and analyze some of its features, from which a few aspects of the general picture emerge.

Special attention is directed to the relationship between the average lifespan of a population and the width of the distribution (variance-mean ratio, also known as the index of dispersion or Fano factor). In the ``Gumbel cases" the mean time to extinction diverges (albeit logarithmically) at the thermodynamic limit, while the width of the distribution remains constant. Therefore, fluctuations become negligible in large systems. This characteristic reflects the negligible effect of demographic noise when the abundance is large.

In certain systems, as we will explore, the average time to extinction is unaffected by the initial population size. In such cases, even in the ``thermodynamic" limit (large initial size), the mean-variance ratio is ${\cal O} (1)$, indicating significant fluctuations. Specifically, we examine a population-genetic model for diploid with dominance and offer insights into the broader scenario.

\subsection{Density-independent dynamics and the Gumbel statistics} \label{sec2a}

We begin with a simple example in which the general answer is attainable and suggest an argument for the general case.

Let us consider a system with no density-dependent effects. In that case, for any single individual the birth and death rates, per unit time, are fixed, i.e., are independent of the state of other individuals. The death rate is taken to be $\mu$ and the birth rate is $\lambda$. If the population is extinction prone, $\mu > \lambda$.

The chance $P_n(t)$ of having $n$ individuals at time $t$ satisfies the following differential equation,
\begin{equation}\label{1}
\frac{d P_n(t)}{d t} = \mu (n+1) P_{n+1} + \lambda (n-1) P_{n-1} - (\mu + \lambda) nP_{n}.
\end{equation}
We would like to solve this equation and to find $P_0$, the chance of extinction, given that $P_n(t=0) = \delta_{n,N_0}$.
To do that we introduce the generating function,
\begin{equation}
G(x,t) = \sum_{n=0}^{\infty} P_n x^n,
\end{equation}
obeying,
\begin{equation}
\dot{G} = \mu \sum_n x^n  (n+1) P_{n+1} + \lambda \sum_n x^n (n-1) P_{n-1} - (\mu + \lambda) \sum_n x^n  nP_{n}.
\end{equation}
Redefinition of indices yields a first order differential equation for $G$,
\begin{equation} \label{4}
\dot{G} = \mu G' + \lambda x^2 G' - (\mu + \lambda) xG' =  [\mu + \lambda x^2- (\mu + \lambda) x]G' = Q(x) G'.
\end{equation}

Eq. (\ref{4}) is a first-order equation that may be solver using characteristics~\cite{kendall1948generalized}.  Every function of the form $G[F(x)+t]$ will solve Eq.  (\ref{4}) if $dF/dx = 1/Q(x)$. For Eq. (\ref{4}) the desired $F$ is,
\begin{equation}
F(x) = \frac{\ln\left(\frac{x-1}{\lambda x-\mu}\right)}{\lambda-\mu}.
\end{equation}

What's left is to determine the functional form of $G[F(x)+t]$, and this has to do with the initial condition. Suppose at $t=0$ we have only one individual. In that case by definition  $G(x,t=0)=x$ and,
\begin{equation} \label{5}
G^{-1}(F(x))=x.
\end{equation}
The solution for Eq. (\ref{5}) is
\begin{equation}
G(F,t=0) = \frac{\mu e^{(\lambda-\mu) F}-1} {\lambda e^{(\lambda-\mu) F}-1}.
\end{equation}
And therefore  the generating function at any time $t$ is,
\begin{equation}
G(F,t) = \frac{\mu e^{(\lambda-\mu) (F+t)}-1} {\lambda e^{(\lambda-\mu) (F+t)}-1}  = \frac{\mu e^{(\lambda-\mu)t}\left(\frac{x-1}{\lambda x-\mu}\right)-1} {\lambda e^{(\lambda-\mu)t}\left(\frac{x-1}{\lambda x-\mu}\right)-1} .
\end{equation}
Hence, the chance that at time $t$ the lineage of a given individual has already gone extinct is
\begin{equation}
Q_{1 \to 0} (t) = 1- \frac{\mu -\lambda }{\mu  e^{t (\mu -\lambda )}-\lambda}.
\end{equation}
Since the dynamics of the lineages of all individuals are statistically identical (no density-dependent effects), if the population at $t=0$ has $N_0$ individuals,
\begin{equation} \label{gumbel0}
Q_{N_0 \to 0} (t) = \left(1- \frac{\mu -\lambda }{\mu  e^{ (\mu -\lambda )t}-\lambda}\right)^{N_0}.
\end{equation}

To see the connection between the distribution (\ref{gumbel0}) and the Gumbel distribution, let us measure time in units of $\mu$, and define a decline parameter $\kappa = 1- (\lambda/\mu)$. When $N_0 \to \infty$, the time $t$ in which all individuals went extinct is large, and therefore
\begin{equation} \label{gumbel1a}
Q_{N_0 \to 0} (t) \approx e^{-\kappa N_0 e^{-\kappa t}}.
\end{equation}
The chance of extinction at $t$ is $P(t) = dQ/dt$.

Now let us define $t = (s+\nu)/\kappa$, where $\nu = \ln [\beta N_0]$ is the point at which the large-$N_0$ distribution of extinction times, $P(t)$, is peaked,  so the second derivative of the cumulative distribution $Q(t)$ vanishes. With that definition,
\begin{equation} \label{gumbel1}
P(s) = e^{-(s+e^{-s})},
\end{equation}
which is the CDF of the Gumbel distribution whose scale parameter is $\beta = 1/\kappa$ and its mode is  $\mu = \nu/\kappa$. The standard deviation of this distribution is $\pi /\sqrt{6 \kappa^2}$, an ${\cal O}(1)$, $N_0$-independent number.

The variance-mean ratio is then,
\begin{equation} \label{VMR}
\rm{VMR} = \frac{\pi^2}{6 \kappa (\nu+\gamma_E)} = \frac{\pi^2}{6 \kappa (\ln N_0+\gamma_E - \ln \kappa)},
\end{equation}
where $\gamma_E$ is Euler's number. Importantly, this ratio decays like $1/\ln N_0$ in the thermodynamic limit.
As explained in Appendix \ref{appA},  the effect of demographic stochasticity is negligible out of the ``extinction zone" in which $n<n_c$. In the region dominated by demographic noise the dynamics is more or less neutral (see section \ref{neutral}), hence the variance of extinction times distribution is proportional to $n_c$. For extinction-prone systems with no density dependence, $n_c$ is $N_0$ independent (see Appendix \ref{appA}).  This feature may change in other scenarios, as demonstrated in the next subsection.

\begin{figure}[h]
	\centering{
		\includegraphics[width=8cm]{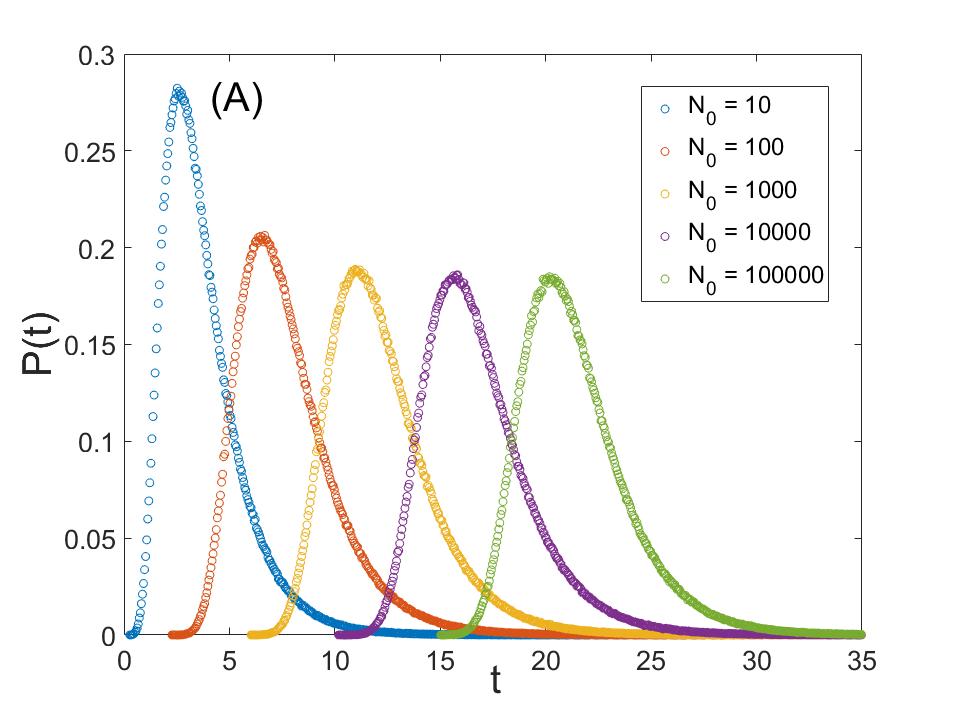}
        \includegraphics[width=8cm]{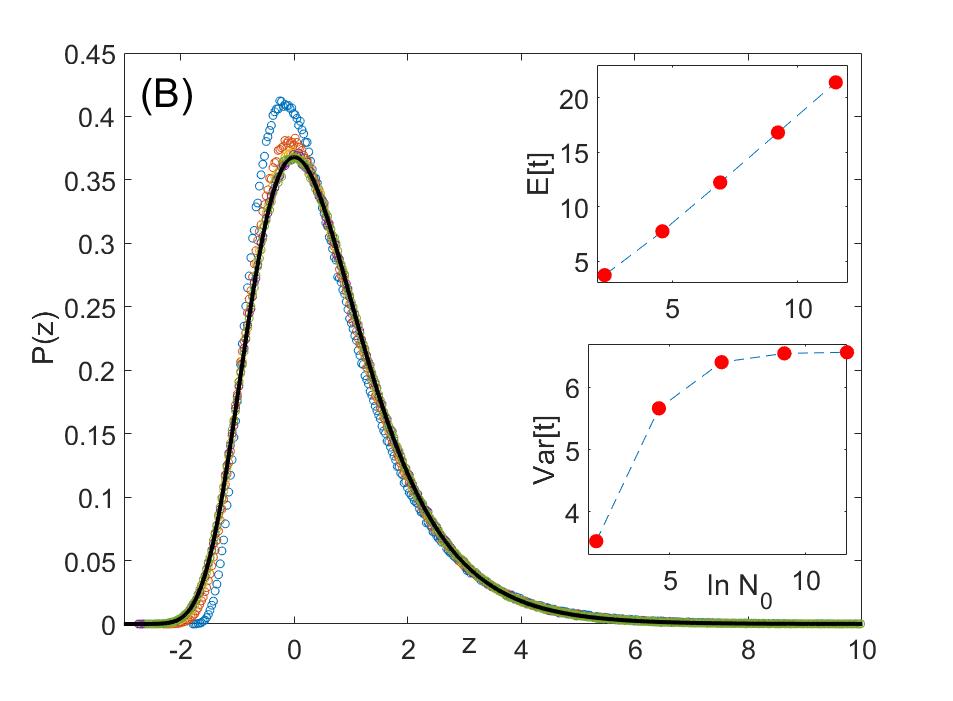}  }

	\caption{The distribution of extinction times, $P(t)$, for an extinction-prone populations when stochasticity is purely demographic (Panel A). The dynamics is logistic, with finite carrying capacity $N_0$. In the presence of $N$ individuals the total death rate is $N$ and the total birth rate is $0.5 N(1-N/N_0)$; the initial population was taken to be $N_0$. The Gumbel distribution parameters $\beta = \sqrt{6{\rm Var[t]}/\pi^2}$ and $\mu = \mathbb{E}[t]-\gamma_E \beta$ ($\gamma_E$ is Euler constant) were extracted for each $N_0$. When a  histogram of the adjusted variable $z = (t-\mu)/\beta$ is plotted (Panel B), all data collapse and fit the Gumbel distribution $\exp(-[z+\exp(-z)])$ (full black line). Small deviations are observed for $N_0 = 10$ and $N_0 = 100$, but above these numbers there is a perfect agreement between the predicted and the observed distribution. The mean and the variance for each $N_0$ are shown in the insets of Panel B. While the mean grows linearly with $\ln N_0$ (upper inset) the variance saturates (lower inset) to its predicted value for $\kappa = 1/2$, namely $2 \pi^2/3 \approx 6.58$.             \label{Gumbel}}
\end{figure}

The general result of ~\citet{hathcock2022asymptotic} may be interpreted as follows. Once the population is in decline, the intraspecific interactions are usually negligible.  The question of extinction time of $N_0$ individuals is thus governed by the chance of the last lineage to go extinct. In the large-$N_0$ limit this becomes the classical extreme-event problem, so as long as the chance of a single lineage to persist decays exponentially at long times, the limit distribution is Gumbel~\cite{fisher1928limiting}. The same answer holds for any other single-lineage distribution which is neither compact nor fat-tailed. The Gumbel statistics is demonstrated, in Figure \ref{Gumbel}, for logistic dynamics with negative growth rate.

\subsection{Density dependent dynamics: Non-Gumbel scenarios}  \label{sec2b}

As pointed out by~\citet{hathcock2022asymptotic}, Gumbel distribution is a universal asymptotic limit of many extinction times statistics provided that the rates of demographic events (transition rates) decrease linearly towards zero at the vicinity of the extinction point. This characteristic reflects the weakening of the interactions between individuals in the extinction zone, so the rate of events is linearly proportional to the number of individuals. When this condition is not fulfilled, the distribution is not Gumbel. In this subsection, we consider a specific example and provide some insights into the more general cases.

As a realistic example, let us consider a population genetics model for diploid with dominance~\cite{haldane1963polymorphism,dean2020stochasticity,karlin1974temporal}. This model describes the dynamics of two alleles, $A$ and $a$, in a randomly mating diploid population. The allele $A$ is always dominant to $a$, so that the phenotype of an $aA$ heterozygote is the same as the phenotype of $AA$.  If the fraction of $a$ alleles in the gamete pool is $x$ and the fraction of $A$ is $(1-x)$, then, after random mating, the zygote genotypes follow classic Hardy-Weinberg proportions, with $AA:Aa:aa$ as $(1-x)^2:2x(1-x):x^2$.

Setting the fitness of $AA$ and $Aa$ phenotype to unity and the fitness of $aa$ to $f<1$, one expects the $a$ allele to disappear from a well-mixed, fixed size population. This purifying selection process is, however, very slow, because an individual will only suffer from low fitness when both of its alleles are of type $a$. Since the number of $a$ homozygotes is proportional to $x^2$, the process is always density-dependent and one expects a non-Gumbel skewed distribution.

\begin{figure}[h]
	\centering{
		\includegraphics[width=8cm]{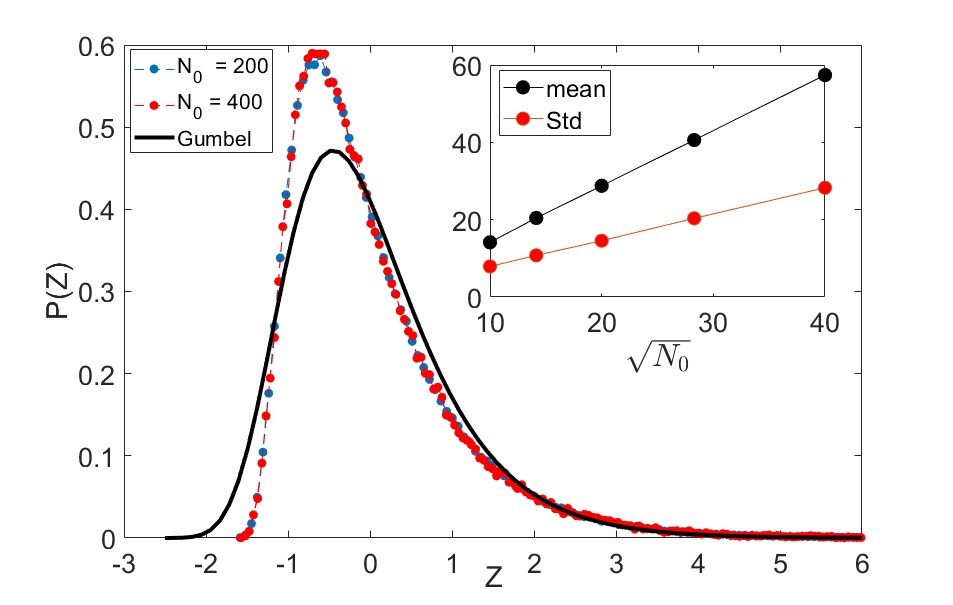}}
	\caption{\textbf{Diploid with dominance}: Main panel: the distribution of normalized extinction times, $P(z)$ vs. $z$ (using the adjusted variable $z = (t-\mu)/\beta$, where $\beta = \sqrt{6{\rm Var[t]}/\pi^2}$ and $\mu = \mathbb{E}[t]-\gamma_E \beta$), where $t$ is the time to extinction of the $a$ allele whose fitness is $f=1/2$. $N$ is the $a$ allele initial frequency, out of total population of $2N_0$ alleles ($N_0$ diploid individuals). Results are shown for $N_0=200$ and for $N_0=400$ (each statistic reflects $10^5$ numerical experiments). Both distributions are almost identical and differ substantially from the Gumbel curve (black line). Inset: the mean (black) and the standard deviation (red) for the same system, plotted vs. $\sqrt{N_0}$ for $N_0 = 100, 200, 400, 800, 1600$. Both quantities scale linearly with $\sqrt{N_0}$, so the variance-mean ratio is finite even in the thermodynamic limit.       \label{diploid}}
\end{figure}

Figure \ref{diploid} shows results from a simulation of this process. In each timestep one individual is chosen to die, so two $a$ alleles are lost with probability $x^2$, one with probability $2x(1-x)$ and the chance of zero $a$ loss is $(1-x)^2$. Then a new individual is introduced, whose two alleles are chosen at random from the gamete pool in which the fraction of $a$ is
\begin{equation}
 \frac{f x^2 + x(1-x)}{f x^2 + 2x(1-x)+(1-x)^2}.
\end{equation}
Although the distribution of the standardized variables is again narrow, and appears to be $N$ independent, it does not satisfy Gumbel statistics, as demonstrated in Figure \ref{diploid}. More importantly, as demonstrated in the inset of figure \ref{diploid}, both the mean and the standard deviation scale with the square root of $N_0$, and therefore the width of the distribution is proportional to its mean even in the thermodynamic limit.

These examples suggest a general insight as to the $N_0$ scaling of the width of the distribution and its mean. As explained in Appendix \ref{appA}, the width reflects the effect of demographic stochasticity, which is relatively weak and becomes prominent only when the deterministic forces are tiny. A population undergoing demographic stochasticity and decline can be described by the Langevin equation:
\begin{equation} \label{eq20}
dn  = -\kappa n^p dt  -\sigma_d \sqrt{n} dW,
\end{equation}
where $\kappa$ is the decay coefficient (related to $\kappa$ and $f$ in the above examples), $p$ is the power that characterizes the interaction between individuals in the dilute limit ($p=1$ for exponential decay with no interactions, $p=2$ for diploid with dominance)  and $\sigma_d$ is the amplitude of demographic variations. The last term in Eq. (\ref{eq20}) becomes important only below $n<n_c$. In Appendix \ref{appA} we show that  $n_c \sim N^{(p-1)/p}$ as long as $p>1$, and $n_c$ is ${\cal O}(1)$ for $p \le 1$. For $n<n_c$ the dynamics is neutral (see section \ref{sec4}), so the contribution of this ``extinction zone" (both to the mean time to extinction and to its standard deviation) is proportional to $n_c$. The regime $n<n_c$ is the only place in which demographic fluctuations are important, so the variance of the extinction time distribution is $n_c^2$.

The mean time to extinction, on the other hand, is the sum of the deterministic timescale, i.e., the time required to decline from $N_0$ to $n_c$, and the stochastic period that scales with $n_c$. The deterministic timescale for the dynamics described by Eq. (\ref{eq20}) is $N_0^{p-1}$ for $p<1$, $\log N_0$ for $p=1$, and is ${\cal O}(1)$ for $p>1$. Accordingly, the variance-mean ratio goes to zero if $p \le 1$ (Assuming both $N_0$ and $N$ diverging). For $p>1$  the mean and the standard deviation both have the same scaling with $n_c \sim N^{(p-1)/p}$, so the variance-mean ratio diverges as $N$ and $N_0$ go to infinity.

\section{Extinction-prone dynamics in stochastic environment} \label{sec3}

In this section we consider the scenario of a population influenced by environmental stochasticity. In what follows, the term "environment" encompasses any external factor that impacts the demographic rates of an \emph{entire} population, including factors such as competition and/or predation pressure from other species. When the environment undergoes stochastic variations, the birth and death rates of the population also fluctuate. Consequently, the overall growth rate (birth rate minus death rate) experiences corresponding variations, leading the population to exhibit either growth or decay. The population is prone to extinction if its mean growth rate is negative~\cite{yahalom2019phase,yahalom2019comprehensive}.

Let us reemphasize the distinction between demographic and environmental stochasticity. The origin of demographic noise (the stochastic characteristics of the birth-death process, as described in the last section) is also the effect of environmental variations on individuals. The distinction between these two forms of stochasticity has to do with their range. When the mean demographic rates remain constant over time and the fluctuations affect individuals in an uncorrelated manner, it is considered demographic noise. On the other hand, if an entire population is affected by the stochasticity, it is classified as environmental stochasticity. Demographic noise is commonly characterized as "white" noise, where different birth or death events are uncorrelated in time. In contrast, the correlation time becomes a significant characteristic of environmental variations.

To wit, let us consider a simple, purely environmental, two-state system (telegraphic noise). We assume that the environment may be in either of two states, say state $1$ and state $2$. The environment remains in a particular state for a certain duration (referred to as the dwell time, which is considered the unit time of the process) before switching to the alternative state with a probability of $1/2$. In each of these states of the environment, the population either grows exponentially or decreases exponentially, so if the number of individuals is $n$,  $\ln n$ increases or decreases linearly with time.

When the number of individuals is large, demographic stochasticity is negligible with respect to environmental stochasticity~\cite{lande2003stochastic}. Therefore, in many studies the effect of demographic stochasticity is taken into account only by introducing a threshold at a given density, below which the population is considered extinct. Recent analyses suggest that this threshold has to be taken at the value of $N$ in which the strength of demographic stochasticity is equal to the strength of environmental stochasticity~\cite{pande2022quantifying,rossberg2022metric}.

Once demographic stochasticity is neglected, the dynamics of $n$ is simply $n(t+\tau) = n(t) \exp(\zeta \tau)$, where $\tau$ is the dwell time and $\zeta$ is the (time dependent) growth exponent (if the environment admits two states, $\zeta$ is either $\zeta_1>0$ or $\zeta_2<0$). Taking $\tau$ as the unit time, one arrives at
\begin{equation}
x_{t+1} =  x_t +\zeta_t,
\end{equation}
where $x = \ln n$. The random walk in $x$-space is characterized by the mean and the variance of  $\zeta$, namely  $\kappa = \overline{\zeta} = (\zeta_1+\zeta_2)/2$, and $\sigma^2 = \rm{Var}[\zeta]$, where $\kappa$, the decline rate, is assumed to be negative.

When the initial population $N_0$ is large, the problem is mapped to the classical first passage time for a biased random walker, as noted a while ago~\cite{lande1988extinction,dennis1991estimation}. Accordingly, if $x_0 = \ln N_0$ is the initial location of the random walker and $x_1 = \ln[N_{th}]$ is the threshold density below which the population is considered extinct, the probability distribution function for the time required to cross the log-space distance $\Delta x = x_0 - x_1 = \ln(N_0/N_{th})$ is given by the inverse Gaussian distribution,
\begin{equation} \label{eq17a}
P(t) = \frac{\Delta x}{\sigma \sqrt{2 \pi t^3}} e^{-\frac{(\Delta x - \kappa t)^2}{2 \sigma^2 t}}.
\end{equation}
The mean of this distribution is $\mathbb{E}[t] = \Delta x/\kappa$ and its variance  ${\rm Var}[t] = \mathbb{E}[t] \sigma^2/\kappa^2$. Therefore, the variance-mean ratio in that case is $N_0$ independent, $\rm{VMR} =\sigma^2/\kappa^2$.

The chance of the system to survive until $t$ (i.e., the cumulative distribution function) is given by
\begin{equation}
Q(t) = \frac{1}{2} \left( 1- {\rm Erf}\left[ \frac{\kappa t-\Delta x}{\sqrt{2 t \sigma^2}}\right]-e^{2\kappa \Delta x/\sigma^2} {\rm Erfc}\left[ \frac{\kappa t+\Delta x}{\sqrt{2 t \sigma^2}}\right]\right).
\end{equation}

In the case of exponential decay ($p=1$) with pure demographic noise considered in section \ref{sec2},  the mean time to extinction is also logarithmic in the initial population size, but the variance and the higher commulants are ${\cal O}(1)$. Here both mean and variance are linear in $\ln N$, so the distribution is much wider than the one that characterizes the purely demographic case. When the noise is demographic, its effect becomes non-negligible only when the number of individuals $n$ is ${\cal O} (1)$ (smaller than $N_{th}$), while for systems with environmental stochasticity the noise affect the system all the way down from $N_0$ to extinction, no matter how large is $N_0$.

The given example focuses on a specific example, namely telegraphic noise. However, at its core, the analysis considers the dynamics of a random walker (in the log-abundance space) with a bias. It can be shown (see, e.g., \cite{yahalom2019comprehensive}, Appendix A) that, as long as the log-abundance steps are not excessively large the diffusion approximation is applicable and the long-term characteristics of the dynamics are solely influenced by the mean and the variance of the $\zeta(t)$ process. Therefore, the results presented above remain valid.

\section{Marginal dynamics with pure demographic stochasticity: the Kimura-Hubbell neutral model} \label{neutral}

In sections \ref{sec2} and \ref{sec3}, we focused on the persistence time statistics of populations prone to extinction. In the upcoming two sections, our aim is to examine the same question but with a focus on marginal populations. These marginal populations are characterized by deterministic dynamics that support a marginally stable manifold, which includes the extinction state. A classic example is the case of competition between two populations or two types that possess identical fitness. For instance, consider two genotypes that differ only by a synonymous mutation, resulting in the same phenotype. In such cases, the system's dynamics become purely stochastic. The famous neutral models proposed by Kimura~\cite{kimura1985neutral,ewens2012mathematical} in population genetics and Hubbell~\cite{Hubbell2001unifiedNeutral,maritan1} in community ecology address such systems, where the dynamics are solely driven by demographic noise.

Under neutral dynamics, species identity is irrelevant. One can consider a single species as a focal species and pool over the effect of all other species together as a single entity (an effective ``rival species").  Therefore, in what follows we examine a single species within a community of $N$ individuals, whose dynamics ends at one of the two absorbing states, i.e., the zero abundance state (extinction) or at abundance $N$ (fixation).

The systems considered in previous sections admit deterministic decline dynamics, so in the long run the overall population never grows beyond its initial value $N_0$, and hence extinction times and extinction statistics are governed by $N_0$, the initial abundance, and not by $N$, the maximum carrying capacity. Under neutral dynamics, a population may either decline to extinction or grow to fixation, and therefore $N$ sets the relevant timescales. $N_0$ affects the statistics only through its relationship with $N$, as explained below.

\subsection{Case I: a macroscopic population}

In Kimura-Hubbel version of the neutral model, with pure demographic stochasticity, one considers the dynamics of $x = n/N$, where $n$ is the number of individuals of a given focal species and $N$ is the total number of individuals. In this subsection we assume that the  initial frequency $N_0/N$ is ${\cal O} (1)$. We would like to obtain the statistics of absorption (either fixation or extinction) times where the dynamics of $P(x,t)$ is given by,
\begin{equation} \label{eq12}
\frac{\partial P(x,\tau)}{\partial \tau} =  \frac{ \partial^2  x (1-x) P(x,\tau)}{\partial t^2}; \qquad P(0,\tau)=P(1,\tau) = 0, \qquad P(x,\tau=0)=\delta(x-1/2).
\end{equation}
Here $\tau$ is the dimensionless timescale $t/N$.

Defining
\begin{equation} \label{eq16}
W(x,\tau) = x(1-x) P(x,\tau),
\end{equation}

 $W$ satisfies,

\begin{equation}
\frac{\partial W(x,\tau)}{\partial \tau} =  x (1-x)  \frac{ \partial^2  W(x,t)}{\partial t^2}.
\end{equation}
Taking $W(x,\tau) = W_m(x)e^{\lambda_m \tau}$, the equation for the eigenfunctions $W_m(x)$ and the eigenvalues $\lambda_m$ is,
\begin{equation} \label{eq11}
 \frac{ \partial^2  W_m(x)}{\partial t^2} - \lambda_m \frac{W_m(x)}{x(1-x)} =0.
\end{equation}

The general solution of (\ref{eq11}) is a linear combination of two independent functions. One is a Meijer G-function that diverges at the origin, so its contribution must vanish (since $P(x)$ vanishes at $x=0$ and at $x=1$, so does $W$). Thus the solution, up to a constant, is given by the other solution, which vanishes at $x=0$,
\begin{equation}
W_m(x) =  x \, _2F_1\left(\frac{1}{2}(1-\sqrt{1-4\lambda_m}),\frac{1}{2}(1+ \sqrt{1-4\lambda_m});2;x\right),
\end{equation}
where $_2F_1$ is the hypergeometric function.

The $\lambda_m$s are determined by the condition $W(x=1)=0$, that yields,
\begin{equation}
W_m(x) = \frac{\cos \left(\frac{1}{2} \pi  \sqrt{1-4 \lambda_m }\right)}{\pi  \lambda_m }
\end{equation}
Therefore,
\begin{equation}
\lambda_m = -m(m+1).
\end{equation}
Since $m$ is an integer, the corresponding eigenfunction simplifies to,
\begin{equation} \label{eq17}
W_m(x) =  x \, _2F_1(-m,m+1;2;x) = x P_m^{(1,-1)}(1-2x)/(m+1),
\end{equation}
where the $P_m^{(\alpha,\beta)}(x)$ are Jacobi polynomials. Accordingly, the general solution to Eq. (\ref{eq11}) takes the form,
\begin{equation} \label{eq13}
W(x,\tau) = \sum_{m=1}^\infty A_m W_m(x) e^{-m(m+1) \tau}.
\end{equation}
The $m=0$ (time-independent) term yields a non-normalizable probability function and therefore it has been discarded.

The constants $A_m$ are determined by the initial condition. The orthogonality relationships of the Jacobi polynomials, when translated to functions of $1-2x$, are
\begin{equation} \label{eq14}
\int_0^1  \frac{x}{1-x} P_m^{(1,-1)}(1-2x) P_n^{(1,-1)}(1-2x) = \delta_{n,m} \frac{(m+1)}{m(2m+1)}.
\end{equation}

To find $A_m$ from $W(x,0)=x(1-x)\delta(x-1/2)$ one multiplies both the left and the right side of this equation by $P_n^{(1,-1)}(1-2x)$, integrates over $x$ from zero to one and applies the relationship (\ref{eq14}). That yields
\begin{equation}
  A_m \frac{1}{m(2m+1)} =
    \begin{cases}
      0 & \text{if m even}\\
    \frac{ P_m^{(1,-1)}(0)}{m+1} =  \frac{(-1)^{m_1} C_{m_1}}{2^m}  & \text{if m odd}
    \end{cases}
\end{equation}
where $m = 2m_1 +1$ and $C_{m_1} = (2m_1)!/(m_1! (m_1+1)!)$ are the Catalan numbers.

\begin{figure}[h]
	\centering{
		\includegraphics[width=7cm]{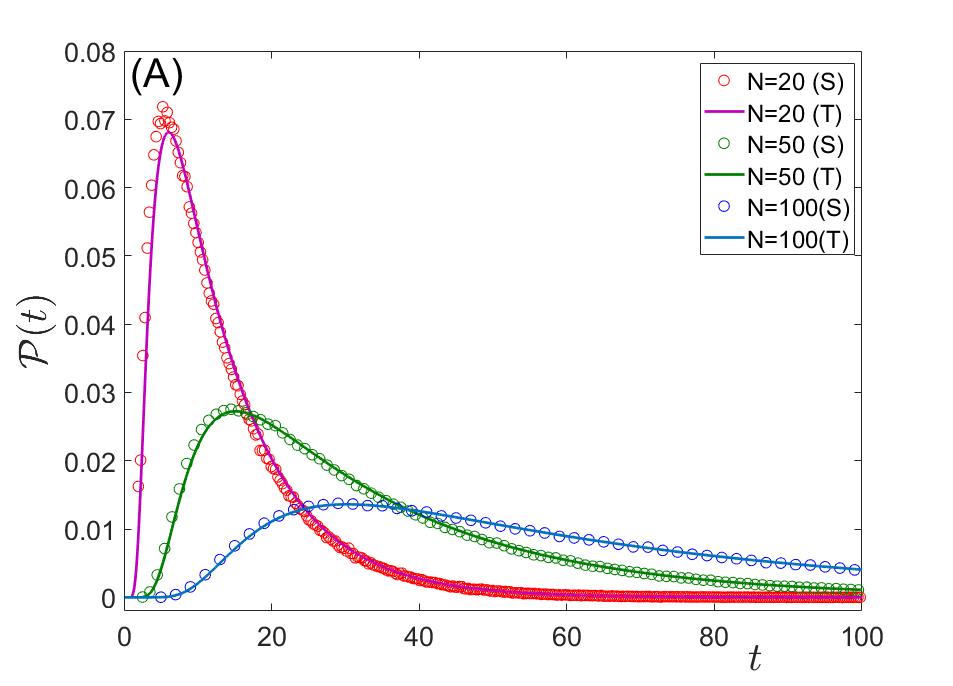}
        \includegraphics[width=7cm]{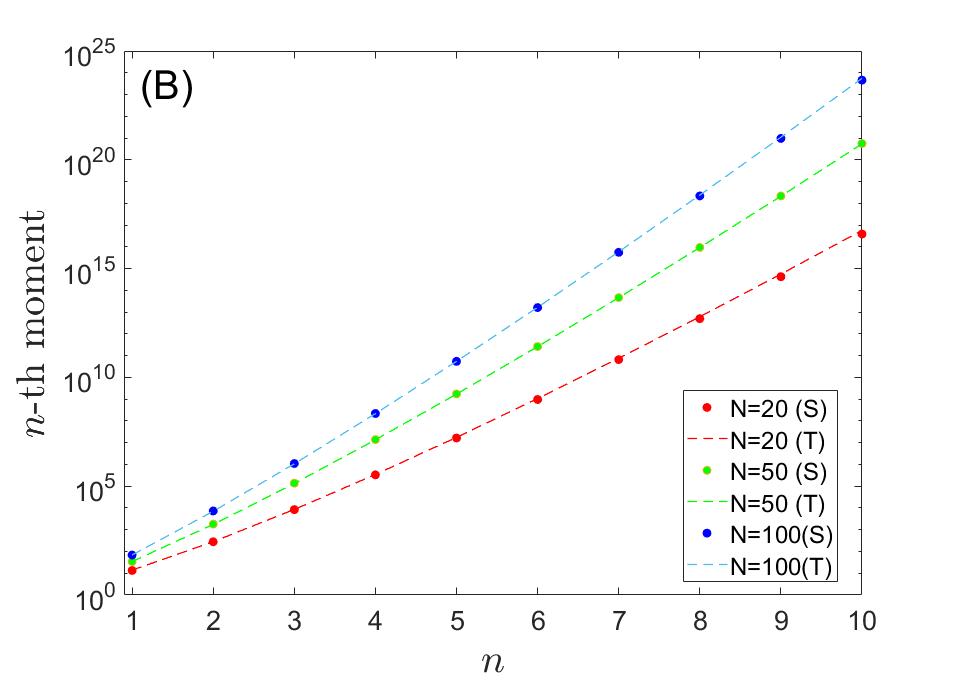}}
	\caption{Panel (A): The extinction probability at $t$, ${\cal P}(t)$, is plotted against $t$ for a population of $N=20,50$ and $100$. The initial condition is $n(t=0)=N/2$, namely $x=1/2$. The result of Eq. (\ref{final}) (full curves) are compared with the normalized distribution obtained numerically (open circles). In the numerical experiment, the chance of the focal population to increase, or to decrease, by one unit in each elementary step is $1/2$, and in each elementary step time is incremented by $1/[2x(1-x)]$. In panel (B) the theoretical prediction for the moments (Eq. \ref{nmom}, dashed lines) are compared with the moments of these distributions (full circles)  \label{fig2}}
\end{figure}

The chance to survive until $t$, $Q(t)$, is given by the integral of $P(x)$ over $x$ from zero to one. Using Eq. (\ref{eq13}), the definition (\ref{eq16}), the relationships between Jacobi polynomials and $W$ and the integral
\begin{equation}
\int_0^1  \ dx \  \frac{ x P_{m}^{(1,-1)}(1-2x)}{x(1-x)} =\frac{2}{m}.
\end{equation}
one finds
\begin{equation}
Q(t) = \int_0^1 dx \   \sum_{m=0}^\infty A_{2m+1} \frac{W_{2m+1}(x)}{x(1-x)} e^{-(2m+1)(2m+2) \tau}  =  \sum_{m=0}^\infty \frac{(-1)^{m+1} C_m}{2^{2m+1}}  (4m+3) e^{-(2m+1)(2m+2)\tau}.
\end{equation}
 Accordingly, the chance of extinction at $\tau$, ${\cal P}(t)$, is
\begin{equation} \label{final}
{\cal P}(t) = -\frac{dQ(t)}{dt} = \frac{1}{N} \sum_{m=0}^\infty \frac{(-1)^m C_m}{2^{2m+1}} (2m+1) (2m+2)  (4m+3) e^{-(2m+1)(2m+2) t/N};
\end{equation}
 Figure \ref{fig2} shows the correspondence between the predicted and the measured ${\cal P}(t) $.

Following the calculation that leads to Eq. (\ref{eq10}) below, one obtains an expression for the asymptotic behavior of the $n$-th moment of the extinction time distribution
\begin{equation} \label{nmom}
\overline{t^n} = B_n N^n n! log(2).
\end{equation}
The number $B_n$ is given by a complex set of hypergeometric functions, however $B_1 = 1$ and in general $B_n \approx \exp(-0.68[n-1])$ provides an excellent approximation for the first $10$ moments, as demonstrated in Figure \ref{fig2}. The mean and the variance are
\begin{equation}
\overline{t} = N \ln 2 \qquad {\rm{Var}[t]} = (2 B_2 -1) (N \ln 2)^2,
\end{equation}
so the VMR scales like $N$.

\subsection{Case II: a single,  neutral mutant}

Let us consider, now, the case of other initial conditions,  $W(x,0) = x(1-x)\delta(x-x_0)$, and  in particular the survival time distribution of a single mutant, $x_0 = 1/N$. Now the general solution for $W(x,\tau)$ takes the form,
\begin{equation}
W(x,\tau) = \sum_{m=1}^\infty \frac{m (2m+1)}{m+1} x_0 P_m^{(1,-1)}(1-2x_0) x P_m(1-2x)  e^{-m(m+1) \tau}.
\end{equation}
Dividing by $x(1-x)$ and integrating over $x$,
\begin{equation}
Q(x,\tau) = \sum_{m=1}^\infty \frac{(2m+1)}{m+1} x_0 P_m^{(1,-1)}(1-2x_0)  e^{-m(m+1) \tau}.
\end{equation}

\begin{figure}[h]
	\centering{
		\includegraphics[width=7cm]{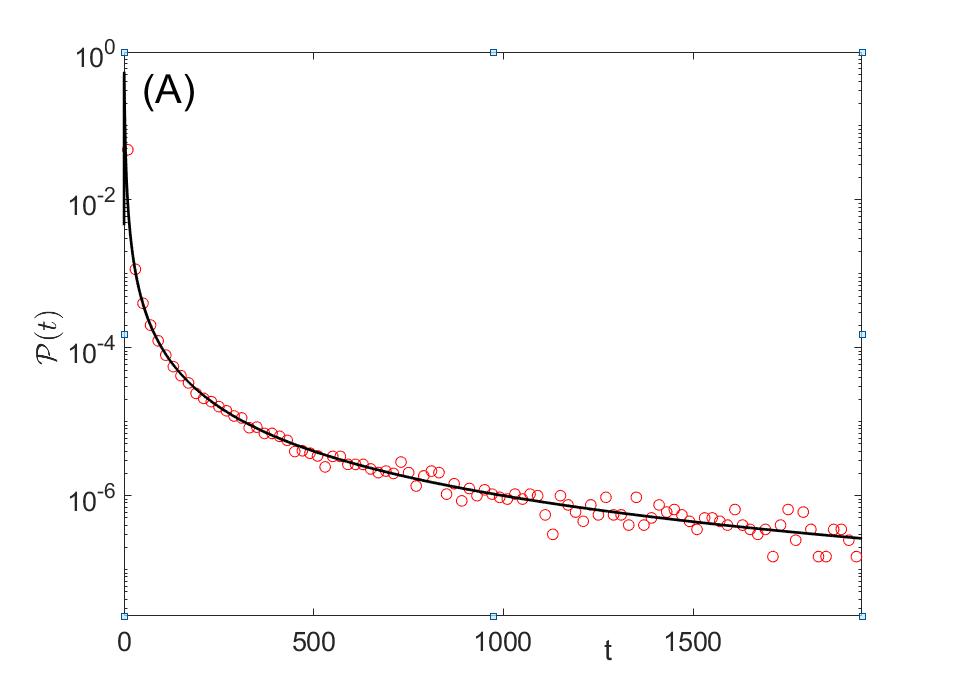}
        \includegraphics[width=7cm]{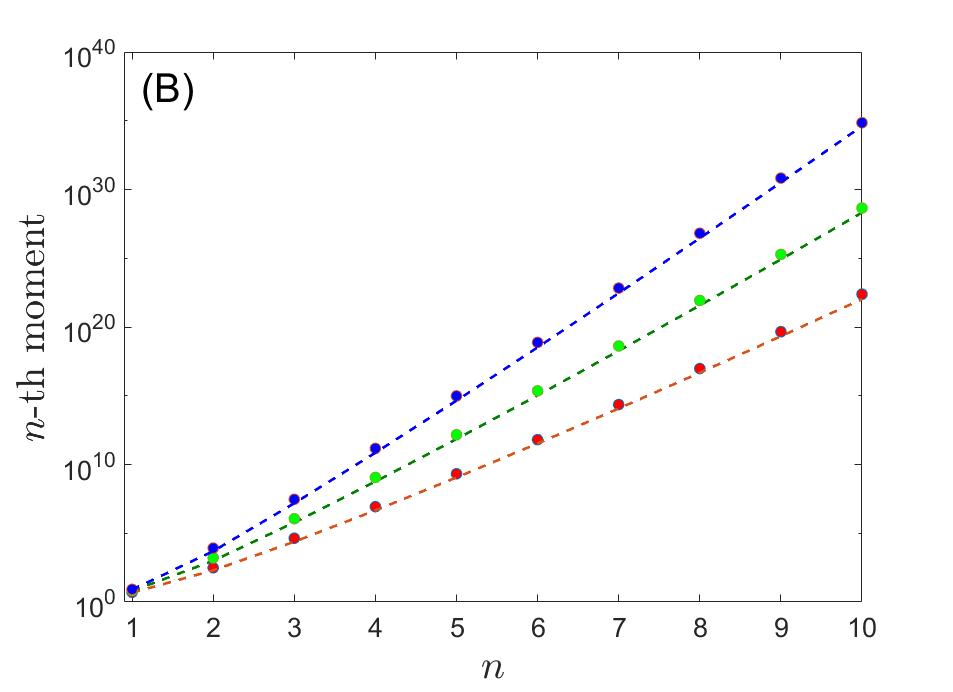}}
	\caption{ Panel (A): The extinction probability of a single mutant ($n(t=0)=1$) at $t$, ${\cal P}(t)$, is plotted against $t$ for a population of $N=100000$.  The result of Eq. (\ref{final_mutant}) (full curve) are compared with the normalized distribution obtained numerically (open circles). In panel (B) the theoretical prediction for the moments (Eq. \ref{final_1}, dashed lines) are compared with the moments of these distributions (full circles)  \label{fig3}}
\end{figure}

If  $N$ is large, for the dynamics of a single mutant ($x_0 =1/N$) one may use the Mehler–Heine formula for the Jacobi polynomials,
\begin{equation} \label{bessel}
P_m^{(1,-1)}(1-2/N) \approx \sqrt{N}J_1 \left( \frac{2m}{\sqrt{N}}\right),
\end{equation}
where $J_1$ is the first Bessel function. Accordingly,
\begin{equation}
Q(x,\tau) \approx \frac{1}{N}  \sum_{m=1}^\infty \frac{(2m+1)}{m+1} J_1 \left( \frac{2m}{\sqrt{N}}\right)  e^{-m(m+1) \tau}.
\end{equation}
Since $J_1$ vanishes at zero, the small-$m$ behavior yields a negligible contribution to the sum. This facilitates the approximation,
\begin{equation}
Q(x,\tau) \approx \frac{2}{N}  \int_{x=1}^\infty  J_1 \left( \frac{2x}{\sqrt{N}}\right)  e^{-x^2 t/N} = 1-e^{-1/t} - \frac{1-e^{-t/N}}{t},
\end{equation}
so the chance of the lineage of a single mutant to reach extinction at $t$ is,
\begin{equation} \label{final_mutant}
{\cal P}(t) = -\frac{dQ(t)}{dt} = \frac{e^{-1/t}}{t^2} - \frac{1-e^{-t/N}(1+t/N)}{t^2}.
\end{equation}
The first moment may be obtained from this expression, and one gets $\overline{t} = \ln N +1 - 2 \gamma_E$, where $\gamma_E$ is Euler's gamma. To get the higher moments we implement the procedure described above,  $\overline{t^n} = -\int t^n [dQ/dt] \ dt$,
\begin{equation} \label{final2}
\overline{t^n} = n! N^{n-1} \sum_{m=1}^{\infty} \frac{2m+1}{m^n (m+1)^{n+1}}P_m^{(1,-1)}(1-2/N).
\end{equation}
Since the main contribution comes from the small-$m$ region, we can approximate $P_m^{(1,-1)}(1-2/N) \approx (m+1)$, and therefore
\begin{equation} \label{final_1}
\overline{t^n} = n! N^{n-1} \sum_{m=1}^{\infty} \frac{2m+1}{m^n (m+1)^n}.
\end{equation}
Figure \ref{fig3} demonstrates the validity of these results.

Note that the time required for a single mutant to be absorbed follows a logarithmic scaling of $\ln N$, whereas the time for a macroscopic population scales linearly with $N$. Additionally, the variance of extinction times for a single mutant is  ${\cal O}(N)$, while for a macroscopic population, it scales with $N^2$. In general, the ratio between the moments described in Equation \ref{final_1} and the corresponding moments in Equation (\ref{nmom}) is an  $N$ factor. This characteristic highlights the fact that an individual either goes extinct within a timescale of ${\cal O}(1)$ or, with a probability that scales like $1/N$, avoids extinction and achieves macroscopic population sizes.

\section{Marginal dynamics with environmental stochasticity: the time-averaged neutral model} \label{sec4}

The neutral model, which we presented in the previous section, was initially introduced by Kimura as a model describing competition between two alleles with equal fitness, and later (with certain modifications) was implemented by Hubbell to describe the dynamics of an ecological community in which all species have equal fitness. Both variations of the model gained immense popularity. In particular, its community ecology version successfully explained the distribution of species abundance in high diversity assemblages using a small number of parameters~\cite{kimura1985neutral,Hubbell2001unifiedNeutral,maritan1,azaele2015towards}.

However, it seems that the neutral model fails to capture the \emph{dynamics} of ecological communities. According to the neutral model, which contains only demographic stochasticity that generates binomial noise, one expects the per-generation changes in abundance to be proportional to the square root of population size. In practice, changes in abundance are usually much larger~\cite{leigh2007neutral}, and usually scale with population size as expected in systems where stochasticity is environmental~\cite{kalyuzhny2014temporal,kalyuzhny2014niche}, not demographic. Similarly, the times to the most recent common ancestor proposed by the neutral model are way too long~\cite{nee2005neutral,ricklefs2006unified}, this phenomenon also reflects the unrealistic "slowness" of neutral dynamics.

To address these issues, the time-averaged neutral model of biodiversity  was proposed~\cite{kalyuzhny2015neutral,danino2018theory,pande2022temporal}. This  is essentially a neutral model with temporal environmental stochasticity. Like the original neutral model, the dynamics is purely stochastic, but in this model, the stochasticity has two sources - both demographic and environmental. All species have the same time-averaged fitness, but at any given moment, there are lower-fitness and higher-fitness species. This immediately leads to abundance variations that scale with population size, as expected, and the theory accounts for both static and dynamic patterns of community assembly~\cite{kalyuzhny2015neutral}.

In the following treatment, we consider a focal species representing a fraction $x$ of the community, competing with another species representing a fraction $1-x$ of the same community. Once again, we address the question of the distribution of times until the focal species reaches either extinction or fixation, this time under environmental noise. If we allow ourselves to neglect the demographic noise, by replacing it with an absorbing boundary condition for populations below a certain threshold, what we obtain is an unbiased random walk in logit ($z=\ln [x/(1-x)]$ space. Therefore, the problem reduces to the distribution of times for a one-dimensional simple random walk with absorbing boundary conditions.

Mathematically speaking, we consider the dynamics of a population whose fraction $x =n/N$ satisfies $\dot{x} = \zeta(t) x(1-x)$, where $\zeta(t)$ is a zero mean stochastic process whose variance is $\sigma^2$. Therefore, the logit variable $z \equiv \ln [x/(1-x)]$ is an unbiased random walk, $z(t) = z_0 \int^t \zeta(t') dt'$. If the threshold fraction $x_{th} = N_{th}/N \ll 1$, the boundary conditions are, to the left $z_{th,L} \approx \ln N_{th}/N$ and to the right  $z_{th,R} \approx  \ln N/N_{th}$. Since there is no bias the absolute value is not important, so we focus on the corresponding diffusion equation
\begin{equation}
\frac{\partial P(z,t)}{\partial t} = D \frac{\partial^2 P(z,t)}{\partial t^2}; \qquad P(0,t)=P(L,t) = 0, \qquad P(x,0)=\delta(z-L/2),
\end{equation}
where $L = z_{th,R}-z_{th,L}$.

The problem is thus equivalent to the heat equation on a $1d$ slab. The general form of the solution is,
\begin{equation}
P(z,t) = \sum_{m=1}^{\infty} A_m \sin\left(\frac{m \pi z}{L} \right) e^{-\lambda_m t}
\end{equation}
where
\begin{equation}
\lambda_m = \frac{D m^2 \pi^2}{L^2}.
\end{equation}
Thus, the solution that satisfies both boundary and initial condition is,
\begin{equation} \label{eq22a}
P(z,t) = \sqrt{\frac{2}{L}} \sum_m (-1)^m \sin\left(\frac{(2m+1) \pi z}{L} \right) e^{-(2m+1)^2 \tau},
\end{equation}
where $\tau \equiv \pi^2 Dt/L^2$.

\begin{figure}[h]
	\centering{
		\includegraphics[width=9cm]{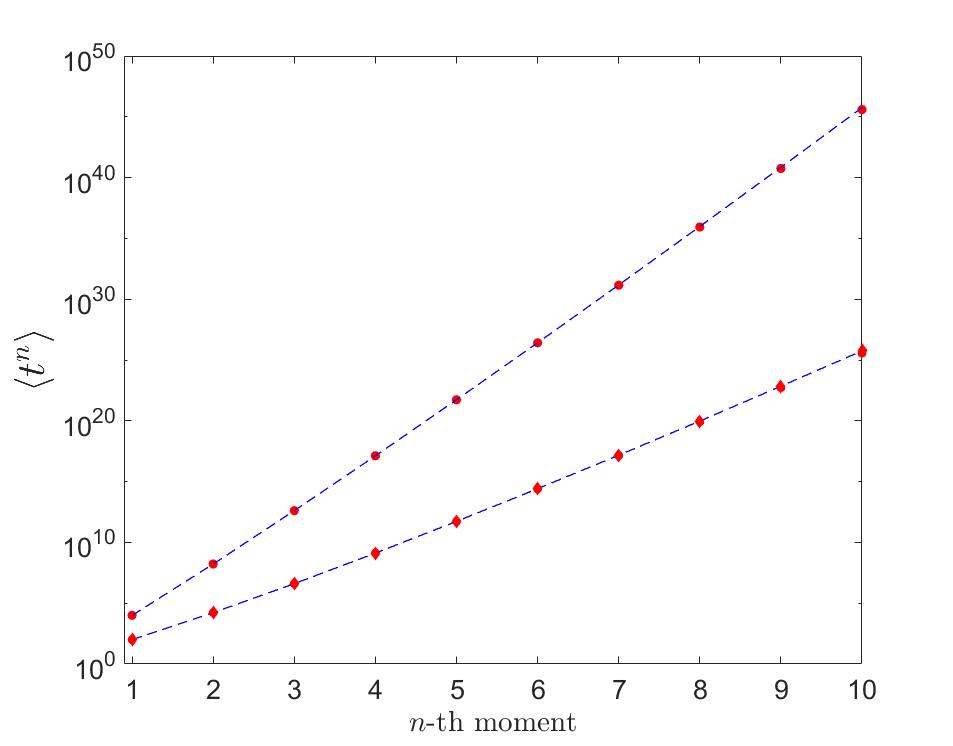}}
	\caption{The $n$-th moment of the extinction time, for a random walker that started at $z=L/2$. In each step the random walker jumps to the left or to the right with probability $1/2$, and time is incremented by one unit. Moments were calculated for extinction times evaluated in $10^5$ numerical experiments for $L=200$ (circles) and $L=20$ (diamonds). Dashed lines are the corresponding predictions from Eq. (\ref{eq10}) with $D=1/2$.    \label{fig1}}
\end{figure}

The chance of the random walker to survive to time $t$, $Q(t)$, is,
\begin{equation} \label{Q1}
Q(t) = \int_0^L P(z,t) dx = \frac{4}{\pi}  \sum_{m } \frac{ (-1)^m}{2m+1} e^{-(2m+1)^2 \tau}.
\end{equation}
The chance of extinction ata  given time $t$ is $-dQ/dt$, and therefore the $n$-th moment of $t$ is given by
\begin{equation}
\overline{t^n} = -\int_0^\infty dt \  t^n \frac{dQ}{dt} = n \int_0^\infty dt t^{n-1} Q(t) = n \left( \frac{L^2}{\pi^2 D} \right)^n   \int_0^\infty  d\tau \tau^{n-1} Q(\tau).
\end{equation}
Evaluating the integral one finds
\begin{equation} \label{eq10}
\overline{t^n} =\frac{ n!}{4^{2n} \pi }  \left( \frac{L^2}{\pi^2 D} \right)^n \left[\zeta(2n+1,1/4)-\zeta(2n+1,3/4) \right],
\end{equation}
where $\zeta$ is the Riemann zeta function.  The agreement between these theoretical predictions and the outcomes of a standard Monte-Carlo simulation is demonstrated in Figure \ref{fig1}.

Here the general scaling of the $n$-th comulant is $L^{2n}$, so the mean time to absorption scales like $\ln^2 N$ and the variance like $\ln^4 N$. As in the case of neutral dynamics with pure demographic stochasticity, the VMR diverges as $N \to \infty$.

For generic initial conditions,  $P(z,0) = \delta(z-z_0)$, Eq. (\ref{Q1}) is replaced by
\begin{equation} \label{Qg}
Q(t) = \int_0^L P(z,t) dz = \frac{4}{\pi}  \sum_{m=0}^{\infty} \frac{\sin[(2m+1)\pi z_0/L}{2m+1} e^{-(2m+1)^2 \tau}.
\end{equation}
so,
\begin{equation} \label{eq15}
\overline{t^n} =\frac{ n!}{4^{2n} \pi }  \left( \frac{L^2}{\pi^2 D} \right)^n \sum_{m=0}^\infty \frac{\sin[(2m+1) \pi z_0/L]}{(2m+1)^{2n+1}}.
\end{equation}
The main contribution to this sum, even for $n=1$, comes from the small-$m$ regime. When $z_0 \to 0$ (close to the absorbing boundaries) the argument of the sine function is negligibly small. Therefore when $z_0 = \epsilon$ the moments are
\begin{equation} \label{eq15a}
\overline{t^n} \approx \frac{4 n!\epsilon}{L}  \left( \frac{L^2}{\pi^2 D} \right)^n (1-4^{-n}) \zeta(2n).
\end{equation}

Again there is a factor of $1/L$ between the ``single mutant" case and the macroscopic population case, because the chance of a single mutant to avoid extinctions on timescales that are ${\cal O}(1)$ and reach macroscopic abundances is proportional to $1/L$.

\begin{figure}[h]
	\centering{
		\includegraphics[width=9cm]{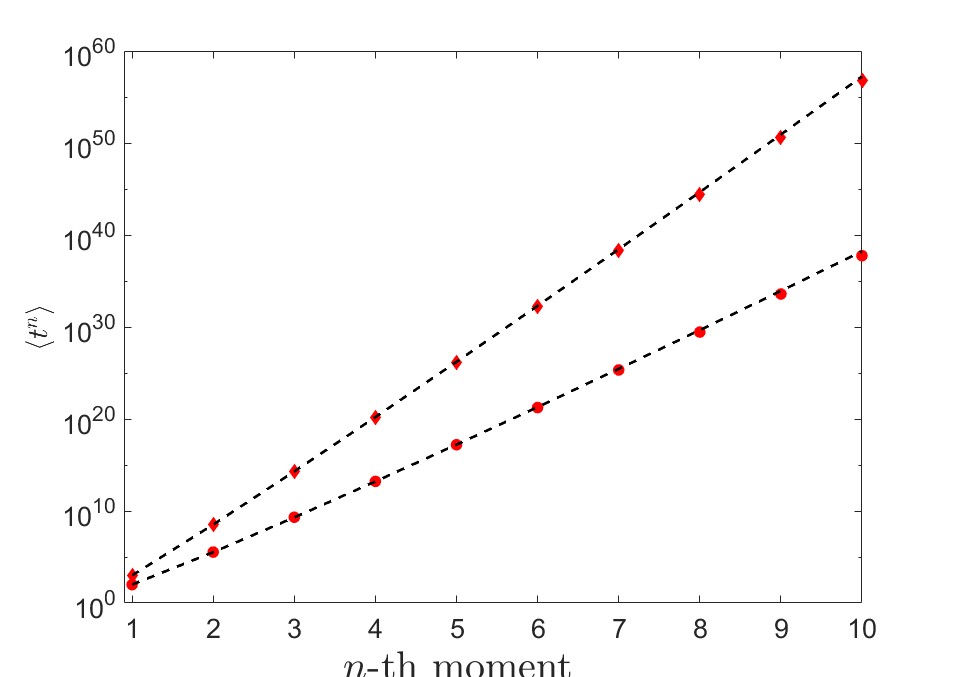}}
	\caption{The $n$-th moment of the extinction time, for a random walker that started at $z_0=1$. In each step the random walker jumps to the left or to the right with probability $1/2$, and time is incremented by one unit. Moments were calculated for extinction times evaluated in $10^5$ numerical experiments for $L=100$ (circles) and $L=1000$ (diamonds). Dashed lines are the corresponding predictions from Eq. (\ref{eq15}) with $D=1/2$.    \label{fig1a}}
\end{figure}

\section{Stable populations}

Now, let's discuss systems that exhibit deterministic dynamics with an attractive fixed point capable of supporting large population. One example is the logistic system described by the equation $dn/dt = rn(1-n/K)$, where $r>0$ (throughout this section, we refer to $K$ as the number of individuals in the equilibrium state). In such cases, the occurrence of extinctions, even in the presence of stochastic fluctuations, is relatively rare. We can think of the stochastic process as a random walk biased towards the equilibrium state. For extinction (or approaching the zero population point) to happen, the random walker would need to take numerous steps "against the current," an event with an extremely low probability.

The stochastic dynamic of a stable system is some sort of a random walk biased away from the extinction point. The path to extinction thus consists of a series of implausible steps, where any plausible step leads to an increase in the population size. Therefore, the most probable decline path is composed of a consecutive sequence of these implausible steps. Under pure demographic stochasticity, this series requires $K$ consecutive death events without any birth event, and the likelihood of this decreases exponentially as $\exp(-c_1 K)$, where $c_1$ is some coefficient. In cases where environmental variations allow for periods of negative growth rate, the most probable path to extinction involves a long period $T$ of adverse weather conditions. The duration $T$ scales logarithmically with $K$, resulting in the frequency of extinctions, which is exponentially rare in $T$, decaying as a power-law function of $K$. These arguments were extensively discussed and presented in detail in~\cite{yahalom2019phase,yahalom2019comprehensive}.

However, beyond the differences in the scaling of the \emph{average} extinction time with $K$, stable systems have a common characteristic that determines the \emph{distribution} around that mean. As mentioned, the extinction event is a rare fluctuation, and the  typical timescale associated with the decline, $T_d \sim \ln K$, is much shorter than the persistence time of a system in the asymptotic limit of large $K$. This separation of timescales, between the decline time and the persistence time, allows us to treat this stochastic process as a binomial process in which, during each increment $T_d$, an extinction event occurs with a tiny probability. If extinction doesn't happen, even if it ``almost" happens (the population declines to small abundance), the system recovers and returns to its equilibrium state. Therefore, the lifetime distribution of stable systems is simply an exponential distribution with an average equal to the average persistence time, as shown in~\cite{yahalom2019comprehensive}.

Mathematically, extensive efforts have been made to calculate the mean time to extinction and determine its numerical value, including the coefficient $c_1$ mentioned earlier or the prefactor of the exponential term~\cite{elgart2004rare,assaf2006spectral,kessler2007extinction,kamenev2008colored}. These studies have revealed that the spectrum of the Markov matrix governing such a stochastic process exhibits several interesting properties. Firstly, it supports an extinction state whose decay rate (log of its eigenvalue) is zero, indicating the absorbing nature of the extinction state. Secondly, there exists a \emph{single} quasi-stationary state whose decay rate decreases to zero as $K$ increases. Finally, the decay rates of all other eigenstates are ${\cal O}(1)$,  independent of $K$. These results are in agreement with the qualitative picture illustrated above: starting from an arbitrary initial state, which is a linear combination of many eigenstates of the corresponding Markov matrix, the system converges to the quasi-stationary state on timescales that are ${\cal O}(1)$, and then the survival probability decays exponentially.

\section{Summary and discussion}

\begin{table} [h]
\noindent%
\begin{tabularx}{\textwidth}{|>{\centering\arraybackslash}X|c|c|c|}
  \hline
  Scenario & Mean  & Variance  & Distribution\\ \hline \hline
  Extinction prone, demographic, density independent ($p=1$) & $\ln N_0$ & ${\cal O}(1)$  & Gumbel (Eq. \ref{gumbel1}) \\  \hline
  Extinction prone, demographic, $p>1$ & $N^{(p-1)/p}$ & $N^{2(p-1)/p}$ & Skewed (Fig \ref{diploid}) \\ \hline
  Extinction prone, environmental & $\ln N_0$ & $\ln N_0$ & Inverse Gaussian (Eq. \ref{eq17a})\\ \hline
  Neutral, demographic, macroscopic population & $N$ & $N^2$ & (Eq. \ref{final}) \\ \hline
  Neutral, demographic, single mutant & $\ln N$ & $N$ &  (Eq. \ref{final_mutant}) \\ \hline
  Neutral + environmental stochasticity, macroscopic population & $\ln^2 N$ &  $\ln^4 N$ & (Eq. \ref{eq22a})  \\
  \hline
  Neutral + environmental stochasticity, single mutant& $\ln N$  &  $\ln^3 N$ & (Eq. \ref{eq15})  \\
  \hline
  Stable population, demographic stochasticity& $\exp(K)$  &  $\exp(2K)$  &  Exponential~\cite{elgart2004rare,assaf2006spectral,kessler2007extinction}  \\
  \hline
  Stable population, environmental stochasticity& $K^\alpha$ (power-law)  &  $K^{2\alpha}$  &  Exponential~\cite{yahalom2019phase,yahalom2019comprehensive}  \\
  \hline
\end{tabularx}
\caption {A summary of the main results.   \label{table1}}

\end{table}

Through this paper we discussed the extinction time statistics in various generic scenarios. The main results we derived or quoted are summarized in Table \ref{table1}. In the mean and in the variance columns of this table we provided only the dependencies of the times on the relevant large parameter, be it the initial population size $N_0$, the total population $N$ or the population at the attractive fixed point $K$.

Perhaps it is worth starting at this point: the determining factor, be it $K$, $N_0$ or $N$. In a stable system, this factor is $K$, the number of focal species individuals' in the stable state. It is independent of the initial population size $N_0$, because the system usually flows towards the stable state. Similarly, it has nothing to do with the total carrying capacity $N$ (how many total individuals, regardless of species, are allowed in the system).

In marginal and neutral systems there is no specific abundance for a particular species. Accordingly, the determining factor is the total carrying capacity of the system, $N$, because every species has a non-negligible chance of reaching it regardless of its initial size. In contrast, in an exponentially decaying system, the initial condition $N_0$ is the only important factor since the population does not generally increase in size.

An exceptional case is when a population undergoes density-dependent extinction dynamics, as demonstrated in the diploid with dominance dynamics. In this case, the deterministic extinction time depends only weakly on the initial population size. Therefore, the factor that governs extinction times is the width of the fluctuations-dominated region, where the system exhibits neutral behavior. Consequently in these cases ($p>1$) the important quantity is again $N$, since it determines the width of the stochastically-dominated zone.

The width of the distribution, and the variance-mean ratio, are governed by the stochastic part of the dynamics. When the origin of these fluctuations is demographic and the deterministic forces take the system to extinction, these fluctuations are important only in a narrow region around zero ($n<n_c$). In the Gumbel case, or by and large when $p \le 1$, this implies that the variance-mean ratio goes to zero in the thermodynamic limit. When $p>1$ two things happen. First, $n_c$ is proportional to $N$, and second, the time required to reach $n_c$, starting from $N_0$, is ${\cal O}(1)$. Therefore, the properties of the distribution of extinction times when $p<1$ are more or less identical to the corresponding properties of a neutral system with $N \sim n_c$.

A significant number of experimental~\cite{drake2006extinction,griffen2008effects,drake2010early} and empirical~\cite{jones1976short,ferraz2003rates,matthies2004population,bertuzzo2011spatial} studies have been dedicated to investigating the distribution of extinction times. However, in order to interpret these results in the context of the archetypal models discussed in this paper, further analysis is required. Nevertheless, we believe that this review article can serve as a point of reference for future analyses of extinction statistics. The key characteristics observed in each study of extinction times, such as their dependence on initial conditions or carrying capacity, first moments,  variance-mean ratio etc., can provide valuable insights for classifying the basic dynamics of the system. This classification can then facilitate more detailed examinations, revealing other, system-specific features. Together, these valuable insights possess the potential to significantly enhance our comprehension of the underlying mechanisms that drive extinctions. Such knowledge can play a pivotal role in bolstering conservation efforts and guiding strategic approaches aimed at safeguarding biodiversity and promoting ecosystem stability.

\noindent
{\bf{Acknowledgments}} We would like to express our gratitude to Stephen P. Ellner for bringing to our attention the experimental works of Drake and his collaborators.

%

\clearpage

\bibliography{extinction_ref}

\begin{thebibliography}{48}%
\makeatletter
\providecommand \@ifxundefined [1]{%
 \@ifx{#1\undefined}
}%
\providecommand \@ifnum [1]{%
 \ifnum #1\expandafter \@firstoftwo
 \else \expandafter \@secondoftwo
 \fi
}%
\providecommand \@ifx [1]{%
 \ifx #1\expandafter \@firstoftwo
 \else \expandafter \@secondoftwo
 \fi
}%
\providecommand \natexlab [1]{#1}%
\providecommand \enquote  [1]{``#1''}%
\providecommand \bibnamefont  [1]{#1}%
\providecommand \bibfnamefont [1]{#1}%
\providecommand \citenamefont [1]{#1}%
\providecommand \href@noop [0]{\@secondoftwo}%
\providecommand \href [0]{\begingroup \@sanitize@url \@href}%
\providecommand \@href[1]{\@@startlink{#1}\@@href}%
\providecommand \@@href[1]{\endgroup#1\@@endlink}%
\providecommand \@sanitize@url [0]{\catcode `\\12\catcode `\$12\catcode
  `\&12\catcode `\#12\catcode `\^12\catcode `\_12\catcode `\%12\relax}%
\providecommand \@@startlink[1]{}%
\providecommand \@@endlink[0]{}%
\providecommand \url  [0]{\begingroup\@sanitize@url \@url }%
\providecommand \@url [1]{\endgroup\@href {#1}{\urlprefix }}%
\providecommand \urlprefix  [0]{URL }%
\providecommand \Eprint [0]{\href }%
\providecommand \doibase [0]{http://dx.doi.org/}%
\providecommand \selectlanguage [0]{\@gobble}%
\providecommand \bibinfo  [0]{\@secondoftwo}%
\providecommand \bibfield  [0]{\@secondoftwo}%
\providecommand \translation [1]{[#1]}%
\providecommand \BibitemOpen [0]{}%
\providecommand \bibitemStop [0]{}%
\providecommand \bibitemNoStop [0]{.\EOS\space}%
\providecommand \EOS [0]{\spacefactor3000\relax}%
\providecommand \BibitemShut  [1]{\csname bibitem#1\endcsname}%
\let\auto@bib@innerbib\@empty
\bibitem [{\citenamefont {Hathcock}\ and\ \citenamefont
  {Strogatz}(2022)}]{hathcock2022asymptotic}%
  \BibitemOpen
  \bibfield  {author} {\bibinfo {author} {\bibfnamefont {D.}~\bibnamefont
  {Hathcock}}\ and\ \bibinfo {author} {\bibfnamefont {S.~H.}\ \bibnamefont
  {Strogatz}},\ }\href@noop {} {\bibfield  {journal} {\bibinfo  {journal}
  {Physical Review Letters}\ }\textbf {\bibinfo {volume} {128}},\ \bibinfo
  {pages} {218301} (\bibinfo {year} {2022})}\BibitemShut {NoStop}%
\bibitem [{\citenamefont {Dornelas}\ \emph {et~al.}(2014)\citenamefont
  {Dornelas}, \citenamefont {Gotelli}, \citenamefont {McGill}, \citenamefont
  {Shimadzu}, \citenamefont {Moyes}, \citenamefont {Sievers},\ and\
  \citenamefont {Magurran}}]{dornelas2014assemblage}%
  \BibitemOpen
  \bibfield  {author} {\bibinfo {author} {\bibfnamefont {M.}~\bibnamefont
  {Dornelas}}, \bibinfo {author} {\bibfnamefont {N.~J.}\ \bibnamefont
  {Gotelli}}, \bibinfo {author} {\bibfnamefont {B.}~\bibnamefont {McGill}},
  \bibinfo {author} {\bibfnamefont {H.}~\bibnamefont {Shimadzu}}, \bibinfo
  {author} {\bibfnamefont {F.}~\bibnamefont {Moyes}}, \bibinfo {author}
  {\bibfnamefont {C.}~\bibnamefont {Sievers}}, \ and\ \bibinfo {author}
  {\bibfnamefont {A.~E.}\ \bibnamefont {Magurran}},\ }\href@noop {} {\bibfield
  {journal} {\bibinfo  {journal} {Science}\ }\textbf {\bibinfo {volume}
  {344}},\ \bibinfo {pages} {296} (\bibinfo {year} {2014})}\BibitemShut
  {NoStop}%
\bibitem [{\citenamefont {Gonzalez}\ \emph {et~al.}(2016)\citenamefont
  {Gonzalez}, \citenamefont {Cardinale}, \citenamefont {Allington},
  \citenamefont {Byrnes}, \citenamefont {Arthur~Endsley}, \citenamefont
  {Brown}, \citenamefont {Hooper}, \citenamefont {Isbell}, \citenamefont
  {O'Connor},\ and\ \citenamefont {Loreau}}]{gonzalez2016estimating}%
  \BibitemOpen
  \bibfield  {author} {\bibinfo {author} {\bibfnamefont {A.}~\bibnamefont
  {Gonzalez}}, \bibinfo {author} {\bibfnamefont {B.~J.}\ \bibnamefont
  {Cardinale}}, \bibinfo {author} {\bibfnamefont {G.~R.}\ \bibnamefont
  {Allington}}, \bibinfo {author} {\bibfnamefont {J.}~\bibnamefont {Byrnes}},
  \bibinfo {author} {\bibfnamefont {K.}~\bibnamefont {Arthur~Endsley}},
  \bibinfo {author} {\bibfnamefont {D.~G.}\ \bibnamefont {Brown}}, \bibinfo
  {author} {\bibfnamefont {D.~U.}\ \bibnamefont {Hooper}}, \bibinfo {author}
  {\bibfnamefont {F.}~\bibnamefont {Isbell}}, \bibinfo {author} {\bibfnamefont
  {M.~I.}\ \bibnamefont {O'Connor}}, \ and\ \bibinfo {author} {\bibfnamefont
  {M.}~\bibnamefont {Loreau}},\ }\href@noop {} {\bibfield  {journal} {\bibinfo
  {journal} {Ecology}\ }\textbf {\bibinfo {volume} {97}},\ \bibinfo {pages}
  {1949} (\bibinfo {year} {2016})}\BibitemShut {NoStop}%
\bibitem [{\citenamefont {Hekstra}\ and\ \citenamefont
  {Leibler}(2012)}]{hekstra2012contingency}%
  \BibitemOpen
  \bibfield  {author} {\bibinfo {author} {\bibfnamefont {D.~R.}\ \bibnamefont
  {Hekstra}}\ and\ \bibinfo {author} {\bibfnamefont {S.}~\bibnamefont
  {Leibler}},\ }\href@noop {} {\bibfield  {journal} {\bibinfo  {journal}
  {Cell}\ }\textbf {\bibinfo {volume} {149}},\ \bibinfo {pages} {1164}
  (\bibinfo {year} {2012})}\BibitemShut {NoStop}%
\bibitem [{\citenamefont {Lande}\ \emph {et~al.}(2003)\citenamefont {Lande},
  \citenamefont {Engen},\ and\ \citenamefont {Saether}}]{lande2003stochastic}%
  \BibitemOpen
  \bibfield  {author} {\bibinfo {author} {\bibfnamefont {R.}~\bibnamefont
  {Lande}}, \bibinfo {author} {\bibfnamefont {S.}~\bibnamefont {Engen}}, \ and\
  \bibinfo {author} {\bibfnamefont {B.-E.}\ \bibnamefont {Saether}},\
  }\href@noop {} {\emph {\bibinfo {title} {Stochastic population dynamics in
  ecology and conservation}}}\ (\bibinfo  {publisher} {Oxford University
  Press},\ \bibinfo {year} {2003})\BibitemShut {NoStop}%
\bibitem [{\citenamefont {Kalyuzhny}\ \emph
  {et~al.}(2014{\natexlab{a}})\citenamefont {Kalyuzhny}, \citenamefont {Seri},
  \citenamefont {Chocron}, \citenamefont {Flather}, \citenamefont {Kadmon},\
  and\ \citenamefont {Shnerb}}]{kalyuzhny2014niche}%
  \BibitemOpen
  \bibfield  {author} {\bibinfo {author} {\bibfnamefont {M.}~\bibnamefont
  {Kalyuzhny}}, \bibinfo {author} {\bibfnamefont {E.}~\bibnamefont {Seri}},
  \bibinfo {author} {\bibfnamefont {R.}~\bibnamefont {Chocron}}, \bibinfo
  {author} {\bibfnamefont {C.~H.}\ \bibnamefont {Flather}}, \bibinfo {author}
  {\bibfnamefont {R.}~\bibnamefont {Kadmon}}, \ and\ \bibinfo {author}
  {\bibfnamefont {N.~M.}\ \bibnamefont {Shnerb}},\ }\href@noop {} {\bibfield
  {journal} {\bibinfo  {journal} {The American Naturalist}\ }\textbf {\bibinfo
  {volume} {184}},\ \bibinfo {pages} {439} (\bibinfo {year}
  {2014}{\natexlab{a}})}\BibitemShut {NoStop}%
\bibitem [{\citenamefont {Kalyuzhny}\ \emph
  {et~al.}(2014{\natexlab{b}})\citenamefont {Kalyuzhny}, \citenamefont
  {Schreiber}, \citenamefont {Chocron}, \citenamefont {Flather}, \citenamefont
  {Kadmon}, \citenamefont {Kessler},\ and\ \citenamefont
  {Shnerb}}]{kalyuzhny2014temporal}%
  \BibitemOpen
  \bibfield  {author} {\bibinfo {author} {\bibfnamefont {M.}~\bibnamefont
  {Kalyuzhny}}, \bibinfo {author} {\bibfnamefont {Y.}~\bibnamefont
  {Schreiber}}, \bibinfo {author} {\bibfnamefont {R.}~\bibnamefont {Chocron}},
  \bibinfo {author} {\bibfnamefont {C.~H.}\ \bibnamefont {Flather}}, \bibinfo
  {author} {\bibfnamefont {R.}~\bibnamefont {Kadmon}}, \bibinfo {author}
  {\bibfnamefont {D.~A.}\ \bibnamefont {Kessler}}, \ and\ \bibinfo {author}
  {\bibfnamefont {N.~M.}\ \bibnamefont {Shnerb}},\ }\href@noop {} {\bibfield
  {journal} {\bibinfo  {journal} {Ecology}\ }\textbf {\bibinfo {volume} {95}},\
  \bibinfo {pages} {1701} (\bibinfo {year} {2014}{\natexlab{b}})}\BibitemShut
  {NoStop}%
\bibitem [{\citenamefont {Chisholm}\ \emph {et~al.}(2014)\citenamefont
  {Chisholm}, \citenamefont {Condit}, \citenamefont {Rahman}, \citenamefont
  {Baker}, \citenamefont {Bunyavejchewin}, \citenamefont {Chen}, \citenamefont
  {Chuyong}, \citenamefont {Dattaraja}, \citenamefont {Davies}, \citenamefont
  {Ewango} \emph {et~al.}}]{chisholm2014temporal}%
  \BibitemOpen
  \bibfield  {author} {\bibinfo {author} {\bibfnamefont {R.~A.}\ \bibnamefont
  {Chisholm}}, \bibinfo {author} {\bibfnamefont {R.}~\bibnamefont {Condit}},
  \bibinfo {author} {\bibfnamefont {K.~A.}\ \bibnamefont {Rahman}}, \bibinfo
  {author} {\bibfnamefont {P.~J.}\ \bibnamefont {Baker}}, \bibinfo {author}
  {\bibfnamefont {S.}~\bibnamefont {Bunyavejchewin}}, \bibinfo {author}
  {\bibfnamefont {Y.-Y.}\ \bibnamefont {Chen}}, \bibinfo {author}
  {\bibfnamefont {G.}~\bibnamefont {Chuyong}}, \bibinfo {author} {\bibfnamefont
  {H.}~\bibnamefont {Dattaraja}}, \bibinfo {author} {\bibfnamefont
  {S.}~\bibnamefont {Davies}}, \bibinfo {author} {\bibfnamefont {C.~E.}\
  \bibnamefont {Ewango}},  \emph {et~al.},\ }\href@noop {} {\bibfield
  {journal} {\bibinfo  {journal} {Ecology letters}\ }\textbf {\bibinfo {volume}
  {17}},\ \bibinfo {pages} {855} (\bibinfo {year} {2014})}\BibitemShut
  {NoStop}%
\bibitem [{\citenamefont {Bergland}\ \emph {et~al.}(2014)\citenamefont
  {Bergland}, \citenamefont {Behrman}, \citenamefont {O'Brien}, \citenamefont
  {Schmidt},\ and\ \citenamefont {Petrov}}]{bergland2014genomic}%
  \BibitemOpen
  \bibfield  {author} {\bibinfo {author} {\bibfnamefont {A.~O.}\ \bibnamefont
  {Bergland}}, \bibinfo {author} {\bibfnamefont {E.~L.}\ \bibnamefont
  {Behrman}}, \bibinfo {author} {\bibfnamefont {K.~R.}\ \bibnamefont
  {O'Brien}}, \bibinfo {author} {\bibfnamefont {P.~S.}\ \bibnamefont
  {Schmidt}}, \ and\ \bibinfo {author} {\bibfnamefont {D.~A.}\ \bibnamefont
  {Petrov}},\ }\href@noop {} {\bibfield  {journal} {\bibinfo  {journal} {PLoS
  Genetics}\ }\textbf {\bibinfo {volume} {10}},\ \bibinfo {pages} {e1004775}
  (\bibinfo {year} {2014})}\BibitemShut {NoStop}%
\bibitem [{\citenamefont {Kalyuzhny}\ \emph {et~al.}(2015)\citenamefont
  {Kalyuzhny}, \citenamefont {Kadmon},\ and\ \citenamefont
  {Shnerb}}]{kalyuzhny2015neutral}%
  \BibitemOpen
  \bibfield  {author} {\bibinfo {author} {\bibfnamefont {M.}~\bibnamefont
  {Kalyuzhny}}, \bibinfo {author} {\bibfnamefont {R.}~\bibnamefont {Kadmon}}, \
  and\ \bibinfo {author} {\bibfnamefont {N.~M.}\ \bibnamefont {Shnerb}},\
  }\href@noop {} {\bibfield  {journal} {\bibinfo  {journal} {Ecology letters}\
  }\textbf {\bibinfo {volume} {18}},\ \bibinfo {pages} {572} (\bibinfo {year}
  {2015})}\BibitemShut {NoStop}%
\bibitem [{\citenamefont {Grilli}(2020)}]{grilli2020macroecological}%
  \BibitemOpen
  \bibfield  {author} {\bibinfo {author} {\bibfnamefont {J.}~\bibnamefont
  {Grilli}},\ }\href@noop {} {\bibfield  {journal} {\bibinfo  {journal} {Nature
  communications}\ }\textbf {\bibinfo {volume} {11}},\ \bibinfo {pages} {1}
  (\bibinfo {year} {2020})}\BibitemShut {NoStop}%
\bibitem [{\citenamefont {Pechenik}\ and\ \citenamefont
  {Levine}(1999)}]{pechenik1999interfacial}%
  \BibitemOpen
  \bibfield  {author} {\bibinfo {author} {\bibfnamefont {L.}~\bibnamefont
  {Pechenik}}\ and\ \bibinfo {author} {\bibfnamefont {H.}~\bibnamefont
  {Levine}},\ }\href@noop {} {\bibfield  {journal} {\bibinfo  {journal}
  {Physical Review E}\ }\textbf {\bibinfo {volume} {59}},\ \bibinfo {pages}
  {3893} (\bibinfo {year} {1999})}\BibitemShut {NoStop}%
\bibitem [{\citenamefont {Dornic}\ \emph {et~al.}(2005)\citenamefont {Dornic},
  \citenamefont {Chat{\'e}},\ and\ \citenamefont
  {Munoz}}]{dornic2005integration}%
  \BibitemOpen
  \bibfield  {author} {\bibinfo {author} {\bibfnamefont {I.}~\bibnamefont
  {Dornic}}, \bibinfo {author} {\bibfnamefont {H.}~\bibnamefont {Chat{\'e}}}, \
  and\ \bibinfo {author} {\bibfnamefont {M.~A.}\ \bibnamefont {Munoz}},\
  }\href@noop {} {\bibfield  {journal} {\bibinfo  {journal} {Physical review
  letters}\ }\textbf {\bibinfo {volume} {94}},\ \bibinfo {pages} {100601}
  (\bibinfo {year} {2005})}\BibitemShut {NoStop}%
\bibitem [{\citenamefont {Assaf}\ and\ \citenamefont
  {Meerson}(2017)}]{assaf2017wkb}%
  \BibitemOpen
  \bibfield  {author} {\bibinfo {author} {\bibfnamefont {M.}~\bibnamefont
  {Assaf}}\ and\ \bibinfo {author} {\bibfnamefont {B.}~\bibnamefont
  {Meerson}},\ }\href@noop {} {\bibfield  {journal} {\bibinfo  {journal}
  {Journal of Physics A: Mathematical and Theoretical}\ }\textbf {\bibinfo
  {volume} {50}},\ \bibinfo {pages} {263001} (\bibinfo {year}
  {2017})}\BibitemShut {NoStop}%
\bibitem [{\citenamefont {Pande}\ \emph {et~al.}(2022)\citenamefont {Pande},
  \citenamefont {Tsubery},\ and\ \citenamefont
  {Shnerb}}]{pande2022quantifying}%
  \BibitemOpen
  \bibfield  {author} {\bibinfo {author} {\bibfnamefont {J.}~\bibnamefont
  {Pande}}, \bibinfo {author} {\bibfnamefont {Y.}~\bibnamefont {Tsubery}}, \
  and\ \bibinfo {author} {\bibfnamefont {N.~M.}\ \bibnamefont {Shnerb}},\
  }\href@noop {} {\bibfield  {journal} {\bibinfo  {journal} {Ecology Letters}\
  }\textbf {\bibinfo {volume} {25}},\ \bibinfo {pages} {1783} (\bibinfo {year}
  {2022})}\BibitemShut {NoStop}%
\bibitem [{\citenamefont {Kimura}(1985)}]{kimura1985neutral}%
  \BibitemOpen
  \bibfield  {author} {\bibinfo {author} {\bibfnamefont {M.}~\bibnamefont
  {Kimura}},\ }\href@noop {} {\emph {\bibinfo {title} {The neutral theory of
  molecular evolution}}}\ (\bibinfo  {publisher} {Cambridge University Press},\
  \bibinfo {year} {1985})\BibitemShut {NoStop}%
\bibitem [{\citenamefont {Hubbell}(2001)}]{Hubbell2001unifiedNeutral}%
  \BibitemOpen
  \bibfield  {author} {\bibinfo {author} {\bibfnamefont {S.~P.}\ \bibnamefont
  {Hubbell}},\ }\href@noop {} {\emph {\bibinfo {title} {The unified neutral
  theory of biodiversity and biogeography}}}\ (\bibinfo  {publisher} {Princeton
  University Press},\ \bibinfo {year} {2001})\BibitemShut {NoStop}%
\bibitem [{\citenamefont {Volkov}\ \emph {et~al.}(2003)\citenamefont {Volkov},
  \citenamefont {Banavar}, \citenamefont {Hubbell},\ and\ \citenamefont
  {Maritan}}]{maritan1}%
  \BibitemOpen
  \bibfield  {author} {\bibinfo {author} {\bibfnamefont {I.}~\bibnamefont
  {Volkov}}, \bibinfo {author} {\bibfnamefont {J.~R.}\ \bibnamefont {Banavar}},
  \bibinfo {author} {\bibfnamefont {S.~P.}\ \bibnamefont {Hubbell}}, \ and\
  \bibinfo {author} {\bibfnamefont {A.}~\bibnamefont {Maritan}},\ }\href@noop
  {} {\bibfield  {journal} {\bibinfo  {journal} {Nature}\ }\textbf {\bibinfo
  {volume} {424}},\ \bibinfo {pages} {1035} (\bibinfo {year}
  {2003})}\BibitemShut {NoStop}%
\bibitem [{\citenamefont {Hathcock}\ and\ \citenamefont
  {Strogatz}(2019)}]{hathcock2019fitness}%
  \BibitemOpen
  \bibfield  {author} {\bibinfo {author} {\bibfnamefont {D.}~\bibnamefont
  {Hathcock}}\ and\ \bibinfo {author} {\bibfnamefont {S.~H.}\ \bibnamefont
  {Strogatz}},\ }\href@noop {} {\bibfield  {journal} {\bibinfo  {journal}
  {Physical Review E}\ }\textbf {\bibinfo {volume} {100}},\ \bibinfo {pages}
  {012408} (\bibinfo {year} {2019})}\BibitemShut {NoStop}%
\bibitem [{\citenamefont {Kendall}(1948)}]{kendall1948generalized}%
  \BibitemOpen
  \bibfield  {author} {\bibinfo {author} {\bibfnamefont {D.~G.}\ \bibnamefont
  {Kendall}},\ }\href@noop {} {\bibfield  {journal} {\bibinfo  {journal} {The
  annals of mathematical statistics}\ }\textbf {\bibinfo {volume} {19}},\
  \bibinfo {pages} {1} (\bibinfo {year} {1948})}\BibitemShut {NoStop}%
\bibitem [{\citenamefont {Fisher}\ and\ \citenamefont
  {Tippett}(1928)}]{fisher1928limiting}%
  \BibitemOpen
  \bibfield  {author} {\bibinfo {author} {\bibfnamefont {R.~A.}\ \bibnamefont
  {Fisher}}\ and\ \bibinfo {author} {\bibfnamefont {L.~H.~C.}\ \bibnamefont
  {Tippett}},\ }in\ \href@noop {} {\emph {\bibinfo {booktitle} {Mathematical
  proceedings of the Cambridge philosophical society}}},\ Vol.~\bibinfo
  {volume} {24}\ (\bibinfo {organization} {Cambridge University Press},\
  \bibinfo {year} {1928})\ pp.\ \bibinfo {pages} {180--190}\BibitemShut
  {NoStop}%
\bibitem [{\citenamefont {Haldane}\ and\ \citenamefont
  {Jayakar}(1963)}]{haldane1963polymorphism}%
  \BibitemOpen
  \bibfield  {author} {\bibinfo {author} {\bibfnamefont {J.}~\bibnamefont
  {Haldane}}\ and\ \bibinfo {author} {\bibfnamefont {S.}~\bibnamefont
  {Jayakar}},\ }\href@noop {} {\bibfield  {journal} {\bibinfo  {journal}
  {Journal of Genetics}\ }\textbf {\bibinfo {volume} {58}},\ \bibinfo {pages}
  {237} (\bibinfo {year} {1963})}\BibitemShut {NoStop}%
\bibitem [{\citenamefont {Dean}\ and\ \citenamefont
  {Shnerb}(2020)}]{dean2020stochasticity}%
  \BibitemOpen
  \bibfield  {author} {\bibinfo {author} {\bibfnamefont {A.}~\bibnamefont
  {Dean}}\ and\ \bibinfo {author} {\bibfnamefont {N.~M.}\ \bibnamefont
  {Shnerb}},\ }\href@noop {} {\bibfield  {journal} {\bibinfo  {journal}
  {Ecology}\ ,\ \bibinfo {pages} {e03098}} (\bibinfo {year}
  {2020})}\BibitemShut {NoStop}%
\bibitem [{\citenamefont {Karlin}\ and\ \citenamefont
  {Levikson}(1974)}]{karlin1974temporal}%
  \BibitemOpen
  \bibfield  {author} {\bibinfo {author} {\bibfnamefont {S.}~\bibnamefont
  {Karlin}}\ and\ \bibinfo {author} {\bibfnamefont {B.}~\bibnamefont
  {Levikson}},\ }\href@noop {} {\bibfield  {journal} {\bibinfo  {journal}
  {Theoretical Population Biology}\ }\textbf {\bibinfo {volume} {6}},\ \bibinfo
  {pages} {383} (\bibinfo {year} {1974})}\BibitemShut {NoStop}%
\bibitem [{\citenamefont {Yahalom}\ and\ \citenamefont
  {Shnerb}(2019)}]{yahalom2019phase}%
  \BibitemOpen
  \bibfield  {author} {\bibinfo {author} {\bibfnamefont {Y.}~\bibnamefont
  {Yahalom}}\ and\ \bibinfo {author} {\bibfnamefont {N.~M.}\ \bibnamefont
  {Shnerb}},\ }\href@noop {} {\bibfield  {journal} {\bibinfo  {journal}
  {Physical review letters}\ }\textbf {\bibinfo {volume} {122}},\ \bibinfo
  {pages} {108102} (\bibinfo {year} {2019})}\BibitemShut {NoStop}%
\bibitem [{\citenamefont {Yahalom}\ \emph {et~al.}(2019)\citenamefont
  {Yahalom}, \citenamefont {Steinmetz},\ and\ \citenamefont
  {Shnerb}}]{yahalom2019comprehensive}%
  \BibitemOpen
  \bibfield  {author} {\bibinfo {author} {\bibfnamefont {Y.}~\bibnamefont
  {Yahalom}}, \bibinfo {author} {\bibfnamefont {B.}~\bibnamefont {Steinmetz}},
  \ and\ \bibinfo {author} {\bibfnamefont {N.~M.}\ \bibnamefont {Shnerb}},\
  }\href@noop {} {\bibfield  {journal} {\bibinfo  {journal} {Physical Review
  E}\ }\textbf {\bibinfo {volume} {99}},\ \bibinfo {pages} {062417} (\bibinfo
  {year} {2019})}\BibitemShut {NoStop}%
\bibitem [{\citenamefont {Rossberg}\ \emph {et~al.}(2022)\citenamefont
  {Rossberg}, \citenamefont {O'Sullivan}, \citenamefont {Terry},\ and\
  \citenamefont {Shnerb}}]{rossberg2022metric}%
  \BibitemOpen
  \bibfield  {author} {\bibinfo {author} {\bibfnamefont {A.}~\bibnamefont
  {Rossberg}}, \bibinfo {author} {\bibfnamefont {J.}~\bibnamefont
  {O'Sullivan}}, \bibinfo {author} {\bibfnamefont {C.}~\bibnamefont {Terry}}, \
  and\ \bibinfo {author} {\bibfnamefont {N.}~\bibnamefont {Shnerb}},\
  }\href@noop {} {\  (\bibinfo {year} {2022})}\BibitemShut {NoStop}%
\bibitem [{\citenamefont {Lande}\ and\ \citenamefont
  {Orzack}(1988)}]{lande1988extinction}%
  \BibitemOpen
  \bibfield  {author} {\bibinfo {author} {\bibfnamefont {R.}~\bibnamefont
  {Lande}}\ and\ \bibinfo {author} {\bibfnamefont {S.~H.}\ \bibnamefont
  {Orzack}},\ }\href@noop {} {\bibfield  {journal} {\bibinfo  {journal}
  {Proceedings of the National Academy of Sciences}\ }\textbf {\bibinfo
  {volume} {85}},\ \bibinfo {pages} {7418} (\bibinfo {year}
  {1988})}\BibitemShut {NoStop}%
\bibitem [{\citenamefont {Dennis}\ \emph {et~al.}(1991)\citenamefont {Dennis},
  \citenamefont {Munholland},\ and\ \citenamefont
  {Scott}}]{dennis1991estimation}%
  \BibitemOpen
  \bibfield  {author} {\bibinfo {author} {\bibfnamefont {B.}~\bibnamefont
  {Dennis}}, \bibinfo {author} {\bibfnamefont {P.~L.}\ \bibnamefont
  {Munholland}}, \ and\ \bibinfo {author} {\bibfnamefont {J.~M.}\ \bibnamefont
  {Scott}},\ }\href@noop {} {\bibfield  {journal} {\bibinfo  {journal}
  {Ecological monographs}\ }\textbf {\bibinfo {volume} {61}},\ \bibinfo {pages}
  {115} (\bibinfo {year} {1991})}\BibitemShut {NoStop}%
\bibitem [{\citenamefont {Ewens}(2012)}]{ewens2012mathematical}%
  \BibitemOpen
  \bibfield  {author} {\bibinfo {author} {\bibfnamefont {W.~J.}\ \bibnamefont
  {Ewens}},\ }\href@noop {} {\emph {\bibinfo {title} {Mathematical population
  genetics 1: theoretical introduction}}},\ Vol.~\bibinfo {volume} {27}\
  (\bibinfo  {publisher} {Springer Science \& Business Media},\ \bibinfo {year}
  {2012})\BibitemShut {NoStop}%
\bibitem [{\citenamefont {Azaele}\ \emph {et~al.}(2015)\citenamefont {Azaele},
  \citenamefont {Maritan}, \citenamefont {Cornell}, \citenamefont {Suweis},
  \citenamefont {Banavar}, \citenamefont {Gabriel},\ and\ \citenamefont
  {Kunin}}]{azaele2015towards}%
  \BibitemOpen
  \bibfield  {author} {\bibinfo {author} {\bibfnamefont {S.}~\bibnamefont
  {Azaele}}, \bibinfo {author} {\bibfnamefont {A.}~\bibnamefont {Maritan}},
  \bibinfo {author} {\bibfnamefont {S.~J.}\ \bibnamefont {Cornell}}, \bibinfo
  {author} {\bibfnamefont {S.}~\bibnamefont {Suweis}}, \bibinfo {author}
  {\bibfnamefont {J.~R.}\ \bibnamefont {Banavar}}, \bibinfo {author}
  {\bibfnamefont {D.}~\bibnamefont {Gabriel}}, \ and\ \bibinfo {author}
  {\bibfnamefont {W.~E.}\ \bibnamefont {Kunin}},\ }\href@noop {} {\bibfield
  {journal} {\bibinfo  {journal} {Methods in Ecology and Evolution}\ }\textbf
  {\bibinfo {volume} {6}},\ \bibinfo {pages} {324} (\bibinfo {year}
  {2015})}\BibitemShut {NoStop}%
\bibitem [{\citenamefont {Leigh}(2007)}]{leigh2007neutral}%
  \BibitemOpen
  \bibfield  {author} {\bibinfo {author} {\bibfnamefont {E.~G.}\ \bibnamefont
  {Leigh}},\ }\href@noop {} {\bibfield  {journal} {\bibinfo  {journal} {Journal
  of Evolutionary Biology}\ }\textbf {\bibinfo {volume} {20}},\ \bibinfo
  {pages} {2075} (\bibinfo {year} {2007})}\BibitemShut {NoStop}%
\bibitem [{\citenamefont {Nee}(2005)}]{nee2005neutral}%
  \BibitemOpen
  \bibfield  {author} {\bibinfo {author} {\bibfnamefont {S.}~\bibnamefont
  {Nee}},\ }\href@noop {} {\bibfield  {journal} {\bibinfo  {journal}
  {Functional Ecology}\ }\textbf {\bibinfo {volume} {19}},\ \bibinfo {pages}
  {173} (\bibinfo {year} {2005})}\BibitemShut {NoStop}%
\bibitem [{\citenamefont {Ricklefs}(2006)}]{ricklefs2006unified}%
  \BibitemOpen
  \bibfield  {author} {\bibinfo {author} {\bibfnamefont {R.~E.}\ \bibnamefont
  {Ricklefs}},\ }\href@noop {} {\bibfield  {journal} {\bibinfo  {journal}
  {Ecology}\ }\textbf {\bibinfo {volume} {87}},\ \bibinfo {pages} {1424}
  (\bibinfo {year} {2006})}\BibitemShut {NoStop}%
\bibitem [{\citenamefont {Danino}\ and\ \citenamefont
  {Shnerb}(2018)}]{danino2018theory}%
  \BibitemOpen
  \bibfield  {author} {\bibinfo {author} {\bibfnamefont {M.}~\bibnamefont
  {Danino}}\ and\ \bibinfo {author} {\bibfnamefont {N.~M.}\ \bibnamefont
  {Shnerb}},\ }\href@noop {} {\bibfield  {journal} {\bibinfo  {journal}
  {Physical Review E}\ }\textbf {\bibinfo {volume} {97}},\ \bibinfo {pages}
  {042406} (\bibinfo {year} {2018})}\BibitemShut {NoStop}%
\bibitem [{\citenamefont {Pande}\ and\ \citenamefont
  {Shnerb}(2022)}]{pande2022temporal}%
  \BibitemOpen
  \bibfield  {author} {\bibinfo {author} {\bibfnamefont {J.}~\bibnamefont
  {Pande}}\ and\ \bibinfo {author} {\bibfnamefont {N.~M.}\ \bibnamefont
  {Shnerb}},\ }\href@noop {} {\bibfield  {journal} {\bibinfo  {journal}
  {Journal of Theoretical Biology}\ }\textbf {\bibinfo {volume} {539}},\
  \bibinfo {pages} {111053} (\bibinfo {year} {2022})}\BibitemShut {NoStop}%
\bibitem [{\citenamefont {Elgart}\ and\ \citenamefont
  {Kamenev}(2004)}]{elgart2004rare}%
  \BibitemOpen
  \bibfield  {author} {\bibinfo {author} {\bibfnamefont {V.}~\bibnamefont
  {Elgart}}\ and\ \bibinfo {author} {\bibfnamefont {A.}~\bibnamefont
  {Kamenev}},\ }\href@noop {} {\bibfield  {journal} {\bibinfo  {journal}
  {Physical Review E}\ }\textbf {\bibinfo {volume} {70}},\ \bibinfo {pages}
  {041106} (\bibinfo {year} {2004})}\BibitemShut {NoStop}%
\bibitem [{\citenamefont {Assaf}\ and\ \citenamefont
  {Meerson}(2006)}]{assaf2006spectral}%
  \BibitemOpen
  \bibfield  {author} {\bibinfo {author} {\bibfnamefont {M.}~\bibnamefont
  {Assaf}}\ and\ \bibinfo {author} {\bibfnamefont {B.}~\bibnamefont
  {Meerson}},\ }\href@noop {} {\bibfield  {journal} {\bibinfo  {journal}
  {Physical review letters}\ }\textbf {\bibinfo {volume} {97}},\ \bibinfo
  {pages} {200602} (\bibinfo {year} {2006})}\BibitemShut {NoStop}%
\bibitem [{\citenamefont {Kessler}\ and\ \citenamefont
  {Shnerb}(2007)}]{kessler2007extinction}%
  \BibitemOpen
  \bibfield  {author} {\bibinfo {author} {\bibfnamefont {D.~A.}\ \bibnamefont
  {Kessler}}\ and\ \bibinfo {author} {\bibfnamefont {N.~M.}\ \bibnamefont
  {Shnerb}},\ }\href@noop {} {\bibfield  {journal} {\bibinfo  {journal}
  {Journal of Statistical Physics}\ }\textbf {\bibinfo {volume} {127}},\
  \bibinfo {pages} {861} (\bibinfo {year} {2007})}\BibitemShut {NoStop}%
\bibitem [{\citenamefont {Kamenev}\ \emph {et~al.}(2008)\citenamefont
  {Kamenev}, \citenamefont {Meerson},\ and\ \citenamefont
  {Shklovskii}}]{kamenev2008colored}%
  \BibitemOpen
  \bibfield  {author} {\bibinfo {author} {\bibfnamefont {A.}~\bibnamefont
  {Kamenev}}, \bibinfo {author} {\bibfnamefont {B.}~\bibnamefont {Meerson}}, \
  and\ \bibinfo {author} {\bibfnamefont {B.}~\bibnamefont {Shklovskii}},\
  }\href@noop {} {\bibfield  {journal} {\bibinfo  {journal} {Physical review
  letters}\ }\textbf {\bibinfo {volume} {101}},\ \bibinfo {pages} {268103}
  (\bibinfo {year} {2008})}\BibitemShut {NoStop}%
\bibitem [{\citenamefont {Drake}(2006)}]{drake2006extinction}%
  \BibitemOpen
  \bibfield  {author} {\bibinfo {author} {\bibfnamefont {J.~M.}\ \bibnamefont
  {Drake}},\ }\href@noop {} {\bibfield  {journal} {\bibinfo  {journal}
  {Ecology}\ }\textbf {\bibinfo {volume} {87}},\ \bibinfo {pages} {2215}
  (\bibinfo {year} {2006})}\BibitemShut {NoStop}%
\bibitem [{\citenamefont {Griffen}\ and\ \citenamefont
  {Drake}(2008)}]{griffen2008effects}%
  \BibitemOpen
  \bibfield  {author} {\bibinfo {author} {\bibfnamefont {B.~D.}\ \bibnamefont
  {Griffen}}\ and\ \bibinfo {author} {\bibfnamefont {J.~M.}\ \bibnamefont
  {Drake}},\ }\href@noop {} {\bibfield  {journal} {\bibinfo  {journal}
  {Proceedings of the Royal Society B: Biological Sciences}\ }\textbf {\bibinfo
  {volume} {275}},\ \bibinfo {pages} {2251} (\bibinfo {year}
  {2008})}\BibitemShut {NoStop}%
\bibitem [{\citenamefont {Drake}\ and\ \citenamefont
  {Griffen}(2010)}]{drake2010early}%
  \BibitemOpen
  \bibfield  {author} {\bibinfo {author} {\bibfnamefont {J.~M.}\ \bibnamefont
  {Drake}}\ and\ \bibinfo {author} {\bibfnamefont {B.~D.}\ \bibnamefont
  {Griffen}},\ }\href@noop {} {\bibfield  {journal} {\bibinfo  {journal}
  {Nature}\ }\textbf {\bibinfo {volume} {467}},\ \bibinfo {pages} {456}
  (\bibinfo {year} {2010})}\BibitemShut {NoStop}%
\bibitem [{\citenamefont {Jones}\ and\ \citenamefont
  {Diamond}(1976)}]{jones1976short}%
  \BibitemOpen
  \bibfield  {author} {\bibinfo {author} {\bibfnamefont {H.~L.}\ \bibnamefont
  {Jones}}\ and\ \bibinfo {author} {\bibfnamefont {J.~M.}\ \bibnamefont
  {Diamond}},\ }\href@noop {} {\bibfield  {journal} {\bibinfo  {journal} {The
  Condor}\ }\textbf {\bibinfo {volume} {78}},\ \bibinfo {pages} {526} (\bibinfo
  {year} {1976})}\BibitemShut {NoStop}%
\bibitem [{\citenamefont {Ferraz}\ \emph {et~al.}(2003)\citenamefont {Ferraz},
  \citenamefont {Russell}, \citenamefont {Stouffer}, \citenamefont
  {Bierregaard~Jr}, \citenamefont {Pimm},\ and\ \citenamefont
  {Lovejoy}}]{ferraz2003rates}%
  \BibitemOpen
  \bibfield  {author} {\bibinfo {author} {\bibfnamefont {G.}~\bibnamefont
  {Ferraz}}, \bibinfo {author} {\bibfnamefont {G.~J.}\ \bibnamefont {Russell}},
  \bibinfo {author} {\bibfnamefont {P.~C.}\ \bibnamefont {Stouffer}}, \bibinfo
  {author} {\bibfnamefont {R.~O.}\ \bibnamefont {Bierregaard~Jr}}, \bibinfo
  {author} {\bibfnamefont {S.~L.}\ \bibnamefont {Pimm}}, \ and\ \bibinfo
  {author} {\bibfnamefont {T.~E.}\ \bibnamefont {Lovejoy}},\ }\href@noop {}
  {\bibfield  {journal} {\bibinfo  {journal} {Proceedings of the National
  Academy of Sciences}\ }\textbf {\bibinfo {volume} {100}},\ \bibinfo {pages}
  {14069} (\bibinfo {year} {2003})}\BibitemShut {NoStop}%
\bibitem [{\citenamefont {Matthies}\ \emph {et~al.}(2004)\citenamefont
  {Matthies}, \citenamefont {Br{\"a}uer}, \citenamefont {Maibom},\ and\
  \citenamefont {Tscharntke}}]{matthies2004population}%
  \BibitemOpen
  \bibfield  {author} {\bibinfo {author} {\bibfnamefont {D.}~\bibnamefont
  {Matthies}}, \bibinfo {author} {\bibfnamefont {I.}~\bibnamefont
  {Br{\"a}uer}}, \bibinfo {author} {\bibfnamefont {W.}~\bibnamefont {Maibom}},
  \ and\ \bibinfo {author} {\bibfnamefont {T.}~\bibnamefont {Tscharntke}},\
  }\href@noop {} {\bibfield  {journal} {\bibinfo  {journal} {Oikos}\ }\textbf
  {\bibinfo {volume} {105}},\ \bibinfo {pages} {481} (\bibinfo {year}
  {2004})}\BibitemShut {NoStop}%
\bibitem [{\citenamefont {Bertuzzo}\ \emph {et~al.}(2011)\citenamefont
  {Bertuzzo}, \citenamefont {Suweis}, \citenamefont {Mari}, \citenamefont
  {Maritan}, \citenamefont {Rodr{\'\i}guez-Iturbe},\ and\ \citenamefont
  {Rinaldo}}]{bertuzzo2011spatial}%
  \BibitemOpen
  \bibfield  {author} {\bibinfo {author} {\bibfnamefont {E.}~\bibnamefont
  {Bertuzzo}}, \bibinfo {author} {\bibfnamefont {S.}~\bibnamefont {Suweis}},
  \bibinfo {author} {\bibfnamefont {L.}~\bibnamefont {Mari}}, \bibinfo {author}
  {\bibfnamefont {A.}~\bibnamefont {Maritan}}, \bibinfo {author} {\bibfnamefont
  {I.}~\bibnamefont {Rodr{\'\i}guez-Iturbe}}, \ and\ \bibinfo {author}
  {\bibfnamefont {A.}~\bibnamefont {Rinaldo}},\ }\href@noop {} {\bibfield
  {journal} {\bibinfo  {journal} {Proceedings of the National Academy of
  Sciences}\ }\textbf {\bibinfo {volume} {108}},\ \bibinfo {pages} {4346}
  (\bibinfo {year} {2011})}\BibitemShut {NoStop}%
\bibitem [{\citenamefont {Danino}\ \emph {et~al.}(2018)\citenamefont {Danino},
  \citenamefont {Kessler},\ and\ \citenamefont {Shnerb}}]{danino2018stability}%
  \BibitemOpen
  \bibfield  {author} {\bibinfo {author} {\bibfnamefont {M.}~\bibnamefont
  {Danino}}, \bibinfo {author} {\bibfnamefont {D.~A.}\ \bibnamefont {Kessler}},
  \ and\ \bibinfo {author} {\bibfnamefont {N.~M.}\ \bibnamefont {Shnerb}},\
  }\href@noop {} {\bibfield  {journal} {\bibinfo  {journal} {Theoretical
  Population Biology}\ }\textbf {\bibinfo {volume} {119}},\ \bibinfo {pages}
  {57} (\bibinfo {year} {2018})}\BibitemShut {NoStop}%
\end{thebibliography}%

\clearpage

\appendix

\section{The width of the extinction zone} \label{appA}

In section \ref{sec2} above we considered the distribution of extinction times for extinction-prone population with demographic stochasticity. In this appendix we present a general argument that allows us to estimate the variance of this distribution.

During the process of extinction, the population is influenced by deterministic forces that drive it towards zero, as well as demographic stochasticity. We can define a critical population size, denoted as $n_c$, above which the deterministic forces dominate, rendering stochasticity negligible. Below $n_c$ the population dynamics is essentially neutral, but the population cannot escape to $n > n_c$ due to the dominant deterministic forces.

In the stochastic regime, once the system reaches $n_c$, both the time to extinction (measured in generations) and its variance scale with $n_c$. Consequently, the time to extinction can be divided into two components: the "deterministic time" required for the population to transition from its initial state to $n_c$, which produces little to no variance, and the "stochastic time" with a mean and variance proportional to $n_c$.

Therefore, the crucial step in getting a semi-quantitative insight regarding the gross features of the extinction-time distribution is to estimate $n_c$. This may be done in a several ways. Here we implement a dominant balance approach to the backward Kolomogorov equation (BKE).

Let us begin with the simplest case of an exponentially decaying population. The BKE, as derived in \cite{danino2018stability}, for example, is
\begin{equation}
T''(x) - \kappa N T' = -\frac{N}{x}.
\end{equation}
This equation was derived for two species competition in a community of $N$ individuals, when $x \ll 1$ is the fraction of the focal species and $T$ is the \textbf{mean} time to extinction. $\kappa$ is the selection parameter, and when $\kappa<0$ the focal species population declines exponentially.

Clearly, the $T'$ term corresponds to the deterministic decline and the  $T''$  term represents stochasticity. If we neglect the stochastic term, $T' = 1/\kappa x$ and therefore $T'' = -1/\kappa x^2$. The stochastic term thus dominate when
\begin{equation}
\frac{1}{\kappa x^2} > \kappa NT' = \frac{N}{x},
\end{equation}
i.e., the stochastic regime is below $x_c = n_c/N = 1/\kappa N$, or $n_c = 1/\kappa$.  Therefore, for large $N_0$ the mean time to extinction scales like $\ln N_0/\kappa$ (this is the timescale required for a population that satisfies $\dot{N} = -\kappa N$ to decline below a certain small value) plus an extinction time that scales like $1/\kappa$ and therefore is negligible when $N_0$ is large. On the other hand the contribution for the variance comes only from the stochastic regime and, following section \ref{sec4}, must scale like $1/\kappa$. These two predictions yield the correct scaling for the parameters $\nu$ and $\beta$ in section \ref{sec2a}.

For the diploid with dominance problem of section \ref{sec2b} the relevant BKE is,
 \begin{equation}
T''(x) - \kappa N x T' = -\frac{N}{2 x}.
\end{equation}
Now $T' \sim 1/\kappa x^2$ and therefore $T'' \sim 1/\kappa x^3$. The first term thus becomes equal to the second term at $x_c = 1/\sqrt{\kappa N}$ so $n_c = \sqrt{N/\kappa}$.

Note that the deterministic time in that case is ${\cal O}(1)$ ($N_0$ independent) and therefore both the mean and the variance scale, in the thermodynamic limit, like $\sqrt{N}$.

Extending this argument one finds that for deterministic dynamics that satisfies $\dot{x} = -\kappa x^p$, the variance scales like $N^{(p-1)/p}$. The deterministic timescale is ${\cal O}(1)$ if $p>1$ and scales like $N_0^{p-1}$ for $p<1$. Note that at $p \to \infty$ the stochastic timescale approaches $N$, since the dynamics becomes neutral.

\end{document}